\documentclass[]{interact}

\usepackage{epstopdf}
\usepackage{subfigure}
\usepackage{natbib}
\usepackage[english]{babel}
\usepackage[utf8]{inputenc}
\usepackage{algorithm}
\usepackage[noend]{algpseudocode}

\usepackage{float}
\usepackage{caption}
\usepackage{hyperref}
\hypersetup{
    colorlinks,
    citecolor=black,
    filecolor=black,
    linkcolor=black,
    urlcolor=black
}
\usepackage{multirow}

\DeclareCaptionLabelSeparator{twospace}{\ ~}
\bibpunct[, ]{(}{)}{;}{a}{}{,}
\renewcommand\bibfont{\fontsize{10}{12}\selectfont}
\theoremstyle{plain}

\theoremstyle{definition}

\theoremstyle{remark}

\newcommand\toolss{20}
\newcommand\datass{12}
\newcommand\placess{98,300}

\begin{document}
	
	 
	\title{How can voting mechanisms improve the robustness and generalizability of toponym disambiguation?}
	
	\author{
\name{Xuke Hu\textsuperscript{a}, Yeran Sun\textsuperscript{b}, Jens Kersten\textsuperscript{a}, Zhiyong Zhou\textsuperscript{c}, Friederike Klan\textsuperscript{a}, Hongchao Fan\textsuperscript{d}}
\affil{\textsuperscript{a}Institute of Data Science, German Aerospace Center,  Germany}
\affil{\textsuperscript{b}Department of Geography, University of Lincoln, UK
}
\affil{\textsuperscript{c}Department of Geography, University of Zurich, Switzerland}
\affil{\textsuperscript{d}Department of Civil and Environmental Engineering, Norwegian University of Science and
Technology, Norway}
}

	\maketitle
	\begin{abstract}
A vast amount of geographic information exists in natural language texts, such as tweets and news. Extracting geographic information from texts is called Geoparsing, which includes two subtasks: toponym recognition and toponym disambiguation, i.e., to identify the geospatial representations of toponyms. This paper focuses on toponym disambiguation, which is usually approached by toponym resolution and entity linking. Recently, many novel approaches have been proposed, especially deep learning-based approaches, such as CamCoder, GENRE, and BLINK. In this paper, a spatial clustering-based voting approach that combines several individual approaches is proposed in order to improve SOTA performance in terms of robustness and generalizability. Experiments are conducted to compare a voting ensemble with \toolss\space latest and commonly-used approaches based on \datass\space public datasets, including several highly ambiguous and challenging datasets (e.g., WikToR and CLDW). The datasets are in six types: tweets, historical documents, news, web pages, scientific articles, and Wikipedia articles, containing in total \placess\space places across the world. The results show that the voting ensemble performs the best on all the datasets, achieving an average \textit{Accuracy@161km} of 0.86, proving the generalizability and robustness of the voting approach. Also, the voting ensemble drastically improves the performance of resolving fine-grained places, i.e., POIs, natural features, and traffic ways. 
	
	\end{abstract}

	
	\section{Introduction}
	\label{Section: Introduction}

\begin{figure}[htbp!]
	  	\centering
	  	\includegraphics[width=14cm]{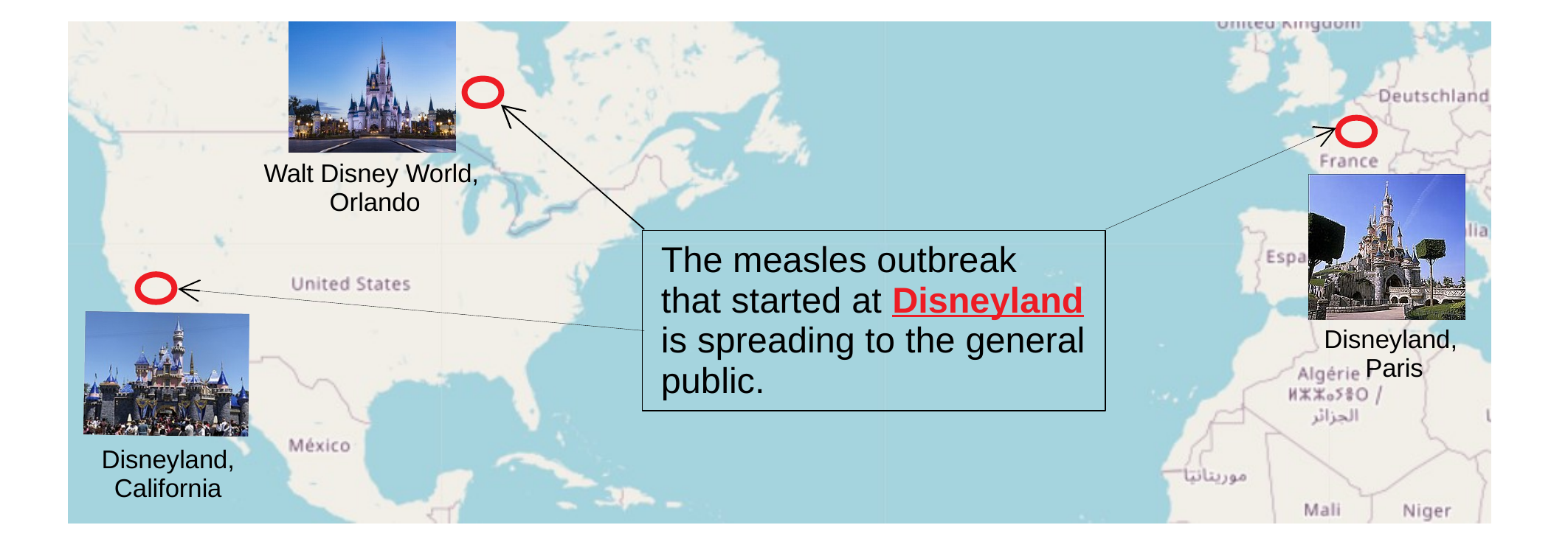}
	  	\caption{Example of toponym disambiguation.\textit{ `Disneyland' } can refer to multiple different locations, such as the park in Paris (France), California (US), Orlando (Canada), and other places named with \textit{ `Disneyland' }.}
	  	\label{disambiguation}
	  \end{figure}

Huge and ever-increasing amounts of semi- and unstructured text data, like news articles, scientific papers, historical archives, and social media posts are available online and offline. These documents often refer to geographic regions or specific places on earth, and therefore contain valuable but hidden geographic information in the form of toponyms or location references. The geographic information is useful not only for scientific studies, such as spatial humanities \citep{gregory2015geoparsing,donaldson2017locating}, but can also contribute to many practical applications \citep{melo2017automated, hu2022location}, such as geographical information retrieval (GIR) \citep{purves2018geographic},  
	 disaster management \citep{shook2016socio}, 
		 disease surveillance \citep{scott2019global},  traffic management \citep{milusheva2021applying}, tourism planning \citep{colladon2019using}, 
 and crime prevention \citep{vomfell2018improving}.
Extracting geographic information from texts is called geoparsing, which includes two subtasks: toponym recognition, i.e., to recognize toponyms from texts, and toponym disambiguation, i.e., to identify their geospatial representations. Toponym recognition has been extensively studied \citep{gazpne2, nLORE, qiuchinesetr, wang2020neurotpr, hu2020GazPNE} and the recognition performance is already very high due to the advancement of deep learning techniques, seeing \citet{hu2022location} for an overview. Therefore, this paper will focus on toponym disambiguation, which is still challenging. Its main task is to remove geo/geo-ambiguities \citep{paradesi2011geotagging}, referring to the situation in which one toponym can refer to more than one geographical location, as shown in Figure \ref{disambiguation}. 

Toponym disambiguation can be approached by entity linking and toponym resolution. Entity linking aims to link an entity (e.g., person, location, event) mentioned in texts to an entry of Knowledge Bases (KBs), such as Wikipedia \citep{wikipedia2004wikipedia}, DBpedia \citep{auer2007dbpedia}, and FreeBase \citep{bollacker2008freebase}, which is the key technology enabling various semantic applications \citep{sevgili2022neural}. 
 Recently, many deep learning-based entity linkers emerged, such as GENRE \citep{decao2020autoregressive}, BLINK \citep{wu2019zero}, LUKE \citep{yamada-etal-2022-global-ed}, Bootleg \citep{orr2020bootleg}, and ExtEnD \citep{barba-etal-2021-extend}, pushing the state-of-the-art performance \citep{sevgili2022neural}. Toponym resolution aims to determine the coordinates of a toponym, focusing only on the location entities, which can be regarded as special cases of entity linking. Quite a few toponym resolution approaches have also been proposed, such as TopoCluster \citep{delozier2015gazetteer},  Edinburgh Geoparser \citep{grover2010use}, CamCoder \citep{gritta2018melbourne}, and CHF \citep{kamalloo2018coherent}. 
Despite the impressive advancement of entity linking techniques, many location entities mentioned in texts cannot be linked to KBs since the current KBs contain only a small proportion of location entries, lacking many small, unpopular, or fine-grained places (e.g., roads and shops). For instance, the largest KB, Wikipedia contains about one million places (i.e., geographically annotated articles), while over 23 million and 12 million places have been recorded in OpenStreetMap \footnote{\url{https://www.openstreetmap.org/}} and GeoNames \footnote{\url{http://www.geonames.org/}}, respectively \citep{hu2022location}. The approaches for toponym resolution normally link toponyms to GeoNames or OpenStreetMap entries. However, according to the results reported in a comparison study \citep{wang2019we}, the performance of the toponym resolution approaches varies by dataset and no one approach can always perform the best.

In this paper, we propose a spatial clustering-based voting approach that combines seven individual approaches to overcome the shortcomings of existing approaches (i.e., entity linking and toponym resolution with the goal of disambiguating toponyms) and further push state-of-the-art performance.
The principle of the proposed voting mechanism is that the minority is subordinate to the dominant (majority). Specifically, we will select the candidate location which receives the most votes from the individual approaches that a voting ensemble combines. Voting approaches have been proven effective \citep{won2018ensemble, hoang2018location, hu2022location} in toponym recognition, the other sub-task of geoparsing. We then compare the voting ensemble with \toolss\space latest and commonly-used entity linking and toponym resolution approaches based on \datass\space public datasets. 

The main contribution of this study is that we propose a more general and robust voting-based approach at the cost of moderate increased computational costs. Furthermore, it is the first time that many competing approaches for toponym disambiguation (especially deep learning-based entity linkers) are compared based on a large number of datasets. Results from this thorough evaluation can help inform future methodological developments for toponym disambiguation, and can help guide the selection of proper approaches based on application needs.

\section{Related works}
\label{related}
There are two main ways to disambiguate toponyms: entity linking and toponym resolution. Entity linking aims at linking a mentioned entity to an entry in an KB, which consists of two steps: named entity recognition and entity disambiguation. It deals with not only locations (toponyms), but also  other types, such as \textit{Person}, \textit{Event}, and \textit{Organization}. 
Recently, many comprehensive surveys \citep{shen2021entity, oliveira2021towards,sevgili2022neural, moller2022survey} on entity linking have been conducted. Therefore, in this section, we will only survey the approaches for toponym resolution by dividing the approaches into three groups based on their basic functional principle:  (1) Rules, (2) Learning and ranking, and (3) Learning and  classification. Note that, many studies, such as \citet{al2017location, ahmed2019real,xu2019dlocrl, yagoub2020newspapers, milusheva2021applying,belcastro2021using,suat2022extraction}, focused only on local events whose geographical scope is known, such as floods or traffic accidents happened in a certain city. Therefore, they would normally match the detected toponyms with the entries in a local gazetteer that contains only the places in the local region to determine the location of the toponyms, without the process of toponym disambiguation.




\subsection{Rules}
Given a target toponym, rule-based approaches first search gazetteers to find all the candidate entries that match or partially match the string of the toponym, and then rank or score the candidate by manually defined IF-THEN rules  \citep{aldana2020adaptive} according to several heuristics or features, such as string similarity, the candidate's population, admin levels, and number of alternative names, spatial distance, and one-sense-per-referent which assumes that all the occurrences of an ambiguous toponym in a text refer to the same location. 
Many rule-based approaches have been proposed, such as Edinburgh Geoparser \citep{grover2010use}, CLAVIN \footnote{\url{https://github.com/Novetta/CLAVIN}}, TwitterTagger \citep{paradesi2011geotagging}, Geoparserpy \citep{middleton2018location}, GeoTxt \citep{karimzadeh2019geotxt}, TAGGS \citep{de2018taggs}, and  CHF, CBH, and SHS \citep{kamalloo2018coherent}. For a toponym in a PubMed article, \citet{weissenbacher2015knowledge} first select the candidate of the toponym, which is in the countries and /or admin level 1 location mentioned in the GenBank \citep{benson2012genbank} records linked to the article. In case an ambiguity cannot be resolved by the above heuristic, the candidate with the highest population will be chosen.
\citet{qi2019thu_ngn} rules that if a toponym appears in  training examples, the candidate with the highest frequency in the training examples is selected. Otherwise, the candidate with the highest population is selected. \citet{de2018taggs} proposed disambiguating toponyms in tweets through grouping tweets. First, for each toponym in an individual tweet, the candidate locations are scored based on different spatial indicators, such as UTC offset, home location of Twitter users, geotags attached to the tweet, and related toponyms. Then, the individual tweets are grouped, and a candidate's score is set to the average score of the candidate of all the individual tweets in the group. The candidate with the highest score will be selected. If multiple candidates have the equal highest score, the one with the highest population is selected. Similarly, for a candidate of a toponym in a tweet, \citet{karimzadeh2019geotxt} accumulates the score made based on nine optional heuristics, such as population, the number of alternate names, GeoNames feature codes, hierarchical relationship, and proximity relationship between two toponyms in the same tweet. 

Implementing rule-based approaches is relatively simple and they are also computational-efficiency. However, manually defined rules are often fragile and ineffective, considering the variability of describing or mentioning toponyms in natural language texts. 

\subsection{Learning and ranking} The workflow of learning and ranking-based approaches is similar to the rule-based approaches. The only difference lies in the rules, which are not explicitly defined but learned from annotated examples. Specifically, a model (or a set of implicit rules) is learned to rank the candidates retrieved from Gazetteers or KBs. 
Many learning and ranking-based  approaches \citep{freire2011metadata, lieberman2012adaptive, santos2015using, xu2019dlocrl,li2019unimelb, wang2019dm_nlp} have been proposed. For example, \citet{lieberman2012adaptive} propose to train a Random Forest model using context-free features (e.g., population and the distance of a candidate to a news's local location) and adaptive context features, such as sibling and proximate relationships between the candidates of the toponyms in a certain context window. The input of the classifiers are the features related to the pair of (\textit{toponym}, \textit{candidate}). 
The output is 1 or 0, indicating if the toponym refers to the candidate or not. Classification confidence is regarded as the ranking score of the candidate.  \citet{li2019unimelb} keep the top 20 candidates researched from GeoNames and then train an SVM model using several features, such as the GeoNames feature codes, name similarity, and the frequency of a candidate in the training set. \citet{wang2019dm_nlp} train a LightGBM \citep{ke2017lightgbm} model using four groups of features: name string similarity, candidate attributes (e.g., popularity), neighboring toponyms, and context features. Context features refer to the contextual similarity between the toponym and the candidate, while the context of the candidate is obtained by requesting the candidate's Wikipage. \citet{santos2015using} applies the LambdaMART learning to rank algorithm \citep{burges2010ranknet} to rank the top 50 most likely candidates retrieved from  Wikipedia using 58 features, such as context features, name similarity, document level features, and geographic features. Rather than disambiguating all the toponyms, \citet{schneider2021portland} apply four weakly supervised machine learning methods (e.g., Logistic Regression, Random Forest, Neural Network, and SVM) to predict whether a mention of \textit{Portland} in a news article refers to \textit{Portland, Oregon} or \textit{Portland, Maine}. Specifically, they use the geotagged articles by NewsStand \citep{teitler2008newsstand} as training data and the frequency of indication words in an article as features. 

Apart from fully supervised approaches, many weakly-supervised and unsupervised approaches have also been proposed to reduce the amount of annotated data required. For example,
\citet{speriosu2013text} proposed WISTR (Wikipedia Indirectly Supervised Toponym Resolver), which generates training examples from geographically annotated Wikipages. Given a page, OpenNLP \footnote{\url{https://opennlp.apache.org/}} is used to detect toponyms on the page, and the true candidate (retrieved from GeoNames) of a toponym is the one closest to the location of the page. A logistic regression model is then trained using features of twenty words at each side of the toponym. \citet{ardanuy2017toponym} propose a weakly supervised approach, named GeoSem. A scoring model is defined to score and rank candidates, which combines and weights multiple features, such as the context similarity between a toponym and a candidate location, geographic closeness to the base location of a collection, and the geographic closeness of a candidate to the candidate of the other toponyms. The parameters (weights) of the model are learned from a small training set. \citet{fize2021deep} proposed disambiguating toponyms based on co-occurrences of toponyms. Specifically, a pair of toponyms, such as  (\textit{'Paris'}, \textit{'Texas'}), with the first as the target toponym and the second as the context features is used as the input of an LSTM network, and the output is the latitude and longitude coordinates of the target toponym. To collect training examples, they use the interlinks between geotagged Wikipedia articles and the inclusion and proximity relationships of toponyms in GeoNames. Rather than training a global model, they train models for different areas, such as France, the US, and Japan. 

Learning and ranking-based approaches can automatically disambiguate toponyms according to multiple features without requiring as much expert knowledge as rule-based approaches do. However, the trained models are often not general enough due to the paucity of sufficient and accurate training data although  unsupervised or weakly supervised techniques have been adopted, making it difficult to use these approaches in many situations \citep{guerini2018toward}.

\subsection{Learning and classification}
Learning and classification-based approaches divide the earth's surface into multiple cells and then locate a toponym to a certain cell. 
For example, \citet{gritta2018melbourne} proposed a CNN-based classification approach, named CamCoder. The input features of CamCoder include the target toponym, the other toponyms in the text, and the context removing the toponyms. CamCoder also uses a geospatial model named Map Vector to produce a vector as the prior probability of candidate locations of the target toponym based on the candidate's population. The prior probability vector is connected with the other three kinds of features for the final classification. 1.4M training examples are generated from over 1M geographically annotated Wikipages.  \citet{kulkarni2020spatial} proposed MLG, a multi-level gazetteer-free geocoding model. The earth's surface is first discretized into cells with S2 geometry from levels 4 (300 km) to 8 (1 km). Then, a CNN model is trained to jointly predict the S2 cells at different levels for a toponym in texts. 
1.76 million training examples are generated from 1.1 M geographically annotated Wikipages.  Different from CamCoder and MLG which use only local context features (e.g., co-occurrence of words and location references in a text), \citet{yan2021integration}  proposed LGGeoCoder, which uses also global context features, including topic embedding and location embedding. Different from the above models that are trained on a large volume of geographically annotated Wikipages, \citet{cardoso2019using, cardoso2021novel} train a Bi-LSTM-based model on training sets of several small datasets, such as LGL and WOTR, and then evaluate the model on corresponding test sets, respectively. \citet{cardoso2021novel} also added 15,000 instances generated from geographically annotated Wikipages to the training sets. 

Some studies also leverage language models (i.e., the spatial distribution of the words in a text) to predict which cell or region is more likely to correspond to the toponym under analysis \citep{speriosu2013text, wing2011simple, delozier2015gazetteer}. This is based on the assumption that apart from toponyms, common language words, such as \textit{`howdy'} and \textit{`phillies'} can often be geographically indicative.  For example, \citet{speriosu2013text} proposed TRIPDL, a toponym resolution approach, built on document geolocators. Specifically,  the earth's surface is first split into 1 degrees by 1 degrees grid cells, and then a probabilistic language model is trained on geographically annotated Wikipages to calculate the probability of locating a document to a cell, using all the words of the document. For a toponym in the document, the cells that contain all the candidates researched from GeoNames are determined, and the candidate whose cell has the largest probability is selected. \citet{delozier2015gazetteer} proposed a gazetteer-independent toponym resolution approach using geographic word profiles, named TopoCluster. The earth's surface is first divided into 60,326 grid cells at a granularity of 0.5 geographic degrees. Then, the per-word spatial distribution is learned based on 700,000 geographically annotated Wikipages. Disambiguation is performed by merging the shared geographic preferences (cells) of a toponym and all words in the context of the toponym. Gazetteer matching can optionally be done by finding the gazetteer entry that matches the target toponym and is closest to the most overlapped cells.

Learning and classification-based approaches are normally trained on geographically annotated Wikipages, which contain around 1 million places at present. However, there are still many places, which are not presented on Wikipedia. For example, there are around 12 million places in GeoNames and 23 million places in OpenStreetMap. Moreover, the text that describes a place in Wikipedia follows a certain pattern, which is different from the other types of texts that mention a place, such as microblogs, news, and scientific articles. 

  \section{Proposed approach}
  In this section, we will introduce the voting approach, summarize the \toolss\space individual approaches that will be used to form or to be compare with a voting ensemble, and illustrate the voting approach with three real examples.

  \subsection{Voting approach}
The idea of this study is inspired by the works of \cite{won2018ensemble, hoang2018location}, which ensemble multiple existing toponym recognition approaches as a voting ensemble, achieving promising recognition performance. Each individual approach has its own limitations while combining multiple individual approaches can overcome these shortcomings. 
Different approaches normally return (or vote for)  different  physical locations (candidates) for a toponym in texts. We count the votes for these physical locations and choose the one with the largest number of votes. Since some approaches outperform the other approaches, the superior approaches' votes should have a higher weight. That is, we will set a higher weight to superiors approaches by copying the coordinate estimation of the approaches multiple times. To realize the voting approach, we adopt DBSCAN \citep{khan2014dbscan}, a density-based clustering approach.
  DBSCAN groups together points that are close to each other based on a distance measurement (eps) and a minimum number of points required to form a group (minPts).
 The workflow of the voting approach is as follows:

\begin{enumerate}
\item Group the coordinate estimation of the individual approaches of a voting ensemble with DBSCAN. The two parameters of DBSCAN are denoted by $eps$ and $minPts$, respectively. 
\item If clusters are formed, select the largest cluster or randomly select one when multiple clusters of same size exist. Treat the centroid of the coordinate estimations in the selected cluster as the voting result.
 \item If no clusters are formed, traverse the individual approaches of the ensemble and treat the first valid estimation as the voting result. 
\end{enumerate}
Invalid estimation by an approach refers to the situation where the approach fails to estimate the coordinates of a toponym, such as the one not appearing in gazetteers. The maximum possible error distance (half of the earth's circumference) is assigned to an invalid estimation, which equals 20,039 km \citep{gritta2020pragmatic}. 


\subsection{Individual approaches}  

Table \ref{inapproach} lists latest and commonly-used approaches for toponym disambiguation. They together cover all types of approaches as discussed in Section \ref{related}. 
Note that, entity linkers would link a detected entity (e.g., toponym) to an entry in a KB, such as Wikipedia and DBpedia. For DBpedia entries, we obtain their coordinates according to the properties of \textit{geo:lat} and \textit{geo:lat} when available. For Wikipedia entries (articles), we obtain their coordinates through an HTTP API \footnote{\url{https://www.mediawiki.org/wiki/API:Geosearch}} if the articles are geographically annotated. When no coordinates can be obtained (e.g., linked to a non-location entry), the coordinates of (0,0) are assigned to target toponyms, denoting an invalid estimation. Details of the \toolss\space approaches are as follows:

 \begin{table} [htbp!]
 \caption{ \toolss\space representative approaches for toponym disambiguation. ML and DL denote traditional machine learning algorithms based on feature engineering and deep learning algorithms, respectively.}
  \label{inapproach}
  \small
\begin{tabular}{lcl}
	
\hline
Name                  & Method Type          \\ \hline
\textbf{DBpedia Spotlight} \citep{mendes2011dbpedia}              & Entity Linker     \\

{\textbf{Entity-Fishing}} \citep{entity-fishing}            & Entity Linker   \\

{ \textbf{MulRel-NEL}} \citep{le2018improving}      & Entity Linker      \\

{\textbf{DCA}} \citep{yang2019learning}            & Entity Linker   \\

{\textbf{BLINK}} \citep{wu2019zero}      & Entity Linker      \\
{\textbf{Bootleg}} \citep{orr2020bootleg}            & Entity Linker   \\

{\textbf{GENRE}} \citep{decao2020autoregressive}            & Entity Linker   \\
{\textbf{ExtEnD}} \citep{barba-etal-2021-extend}            & Entity Linker   \\
{\textbf{LUKE}} \citep{yamada-etal-2022-global-ed}            & Entity Linker   \\

{\textbf{Nominatim}} \footnote{\url{https://nominatim.org/}}           & Geocoder       \\
\textbf{Adaptive learning} \citep{lieberman2012adaptive}   & ML (Ranking)  \\

\textbf{Edinburgh Geoparser} \citep{grover2010use} & Rule     \\
\textbf{Population-Heuristics} \citep{speriosu2013text}   & Rule  \\

\textbf{CLAVIN} \footnote{\url{https://github.com/Novetta/CLAVIN}}               & Rule     \\
\textbf{TopoCluster} \citep{delozier2015gazetteer}        & ML (Classification) \\
\textbf{Mordecai} \citep{halterman2017mordecai}           & Rule     \\
\textbf{CBH, SHS, CHF} \citep{kamalloo2018coherent}                & Rule  \\
\textbf{CamCoder} \citep{gritta2018melbourne}            & DL (Classification)  \\ \hline
\end{tabular}
\end{table}

\begin{itemize}

    

    \item \textbf{DBpedia Spotlight} is a popular entity linking tool. We use the provided HTTPS API  \footnote{\url{https://www.dbpedia-spotlight.org/api}} to annotate and link entities in texts. We treat the entities whose corresponding DBpedia entry has the properties of \textit{geo:lat} and \textit{geo:lat} as toponyms. We will compare DBpedia Spotlight with the other approaches only on the correctly recognized toponyms by DBpedia Spotlight, which is a subset of gold toponyms.
    
    \item \textbf{Entity-Fishing} is an entity linker based on Random Forest and Gradient Tree Boosting. We use the spaCy wrapper  \footnote{\url{https://github.com/Lucaterre/spacyfishing}} of Entity-Fishing, and modify the code to input gold toponyms to the entity disambiguation step. 

\item \textbf{MulRel-NEL} is a neural entity-linking approach that links a mention to Wikipedia. We use the provided API \footnote{\url{https://github.com/informagi/REL}} of  Radboud Entity Linker (REL) \citep{vanHulst:2020:REL}, which uses \textbf{mulrel-nel} for entity disambiguation. We input gold toponyms to the entity disambiguation step.


\item \textbf{DCA} is a neural entity-linking approach that links a mention to Wikipedia. We retrained the model based on the public and widely used AIDA CoNLL-YAGo dataset \citep{hoffart2011robust}. We then modify the code \footnote{\url{https://github.com/YoungXiyuan/DCA}} to input gold toponyms to the entity disambiguation step.

\item \textbf{Bootleg} is a self-supervised named entity disambiguation approach, using a transformer architecture. We use the provided model which is trained on weakly-labeled training data (1.7 times of the originally labeled entities in Wikipedia), and then modify the code \footnote{\url{https://github.com/HazyResearch/bootleg}} to input gold toponyms to the model.

 \item \textbf{BLINK} is an  entity linker based on fine-tuned BERT \citep{devlin2018bert}. We use the provided model that was first pre-trained on nearly 9M unique triples document-mention-entity from Wikipedia and then fine-tuned on the TACKBP-2010 training dataset \citep{ji2010overview}. We modify the code \footnote{\url{https://github.com/facebookresearch/BLINK}} to input gold toponyms to the entity disambiguation step. BLINK returns a list of candidate Wikipedia entities ranked by prediction probability. We choose the highest ranked candidate (Wikipedia article), which is geographically annotated  as the estimation. 

     \item \textbf{GENRE} (Generative ENtity REtrieval) uses a transformer-based architecture. GENRE was first pre-trained on nearly 9M unique triples document-mention-entity from Wikipedia and then fine-tuned with the AIDA dataset. We use the provided model and  code \footnote{\url{https://github.com/facebookresearch/GENRE}} and input gold toponyms to the entity disambiguation step. GENRE returns a list of candidate Wikipedia entities ranked by prediction probabilities. We choose the highest ranked candidate (Wikipedia article), which is geographically annotated  as the estimation. 

   \item \textbf{LUKE} is a global entity disambiguation model based on BERT. It uses both local (word-based) and global (entity-based) contextual information and was trained on a large entity-annotated corpus generated from Wikipedia. We use the trained model and modify the code \footnote{\url{https://github.com/studio-ousia/luke/tree/master/examples/entity_disambiguation}} to input gold toponyms to the model. 
     
\item \textbf{ExtEnD} (Extractive Entity Disambiguation) adopts Transformer-based architectures. ExtEnD was first pretrained on the same Wikipedia dataset as BLINK and then fine-tuned on the AIDA dataset. We use the fine-tuned model and modify the code \footnote{\url{https://github.com/SapienzaNLP/extend}} to input gold toponyms to the model. 

         \item \textbf{Nominatim} is a geocoder, built on  OpenStreetMap data. It cannot disambiguate a toponym. Instead, it returns a list of candidate locations on earth by name (toponym) and address. We keep the first one it returns. Nominatim is used as a baseline system in this study.

        \item \textbf{Population-Heuristics} uses the heuristic of the largest population to disambiguate toponyms. We implement the approach by searching for all the candidates of a toponym from GeoNames and selecting the one with the largest population. It is also used as a baseline system in this study.

    \item \textbf{Edinburgh Geoparser}  \footnote{\url{https://www.ltg.ed.ac.uk/software/geoparser/}} is a geoparsing tool developed by the Language Technology Group (LTG) at the Edinburgh University. Users can select one of the two gazetteers, GeoNames and Unloc, for toponym resolution. Both of them are accessed using a service hosted by the University of Edinburgh Information Services. We will compare Edinburgh Geoparser with the other approaches only on the correctly recognized toponyms by Edinburgh Geoparser, which is a subset of gold toponyms.

\item \textbf{CLAVIN} Cartographic Location And Vicinity INdexer (CLAVIN) applies Apache OpenNLP for toponym recognition and several heuristics and fuzzy search for toponym resolution. We modify its implementation to input gold toponyms to the toponym resolution step. 

\item \textbf{Adaptive Learning} is a Random Forest-based toponym resolution approach. We use its implementation \footnote{\url{https://github.com/ehsk/CHF-TopoResolver}} to retrain a model based on one dataset, i.e., LGL \citep{lieberman2010geotagging}, and apply the model to disambiguate toponyms in the other datasets. We modify its implementation to input gold toponyms to the toponym resolution step.

  \item \textbf{Mordecai} is a geoparsing tool, which uses word2vec for inferring the correct country for a set of toponyms in texts. We modify its implementation \footnote{\url{https://github.com/openeventdata/mordecai}} to input gold toponyms to the toponym resolution step. 
  
      \item \textbf{CBH,SHS,CHF} are three rule-based approaches proposed by \cite{kamalloo2018coherent}. Context-Bound Hypothese (CBH) utilizes the \textit{geo-center inheritance hypothesis} and \textit{near-location hypothesis}. Spatial-Hierarchy Set (SHS) utilizes the containment and sibling relationships among toponyms in a document. Context-Hierarchy Fusion (CHF) fuses SHS and CBH. We modify their implementation \footnote{\url{https://github.com/ehsk/CHF-TopoResolver}} to input gold toponyms to the toponym resolution step.



    \item \textbf{TopoCluster}  is a language model-based geoparsing tool. We modify its implementation \footnote{\url{https://github.com/grantdelozier/TopoCluster}} to input gold toponyms to the toponym resolution step.

    \item \textbf{CamCoder} is a CNN-based geoparsing tool, which was trained on nearly 1M geographically annotated Wikipedia articles. We use the trained model and modify its implementation \footnote{\url{https://github.com/milangritta/Geocoding-with-Map-Vector}} to input gold toponyms to the toponym resolution step.
\end{itemize}

\subsection{Examples}
	   	 \begin{figure} [h!] 
	   	 \small
	  	\centering
	  	\includegraphics[width=12cm]{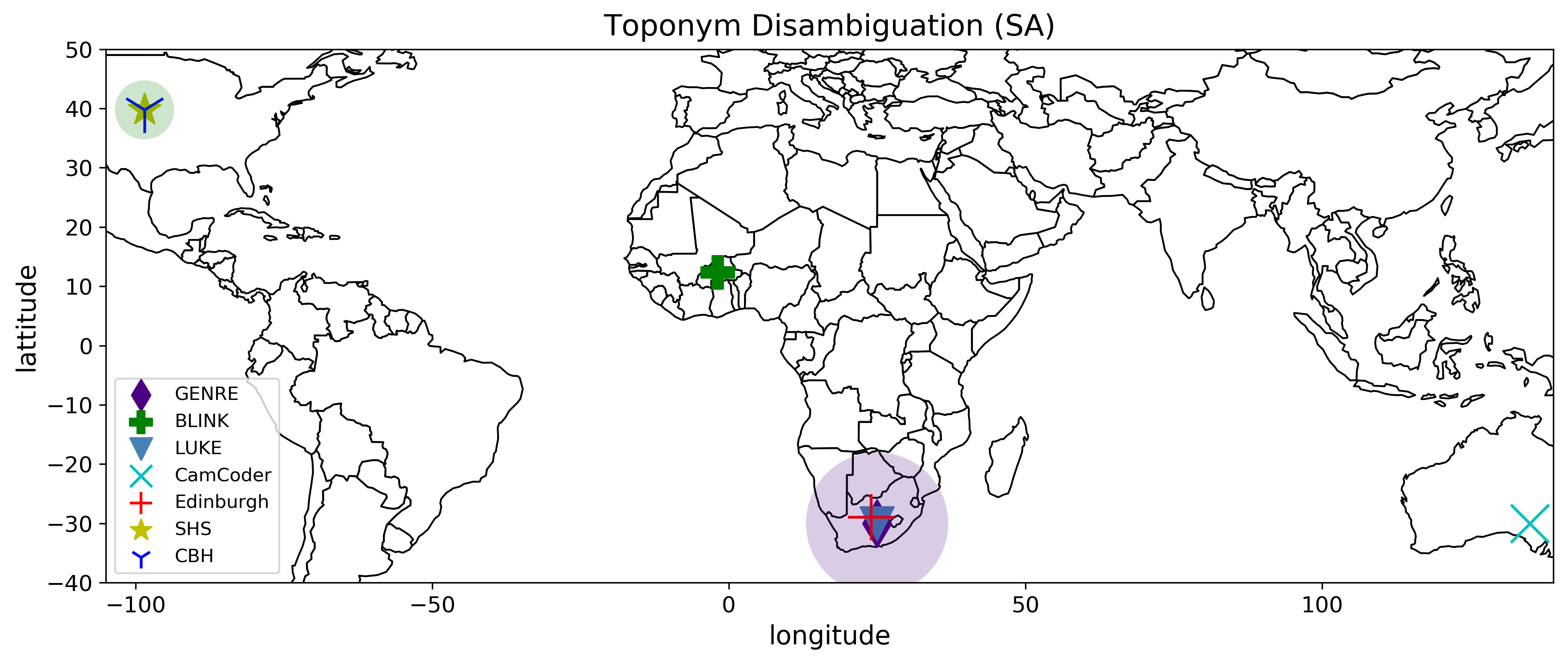}
	  	\caption{An example to show how the voting approach works. The target toponym is \textit{‘SA’}, whose true location is in the largest cluster (purple circle). The context of the toponym is: \textit{‘Lack of transparency and accountability, plus serious allegations of corruption have been leveled against chiefs. For instance, Kgosi (chief) Nyalala Pilane of the Bakgatla-ba-Kgafela community — perhaps even more than any other chief in \textbf{SA} — has been the subject of a litany of maladministration and corruption allegations’}.}
	  	\label{example1}
	  \end{figure}

 	   	 \begin{figure}[h!]
	   	 \small
	  	\centering
	  	\includegraphics[width=8cm]{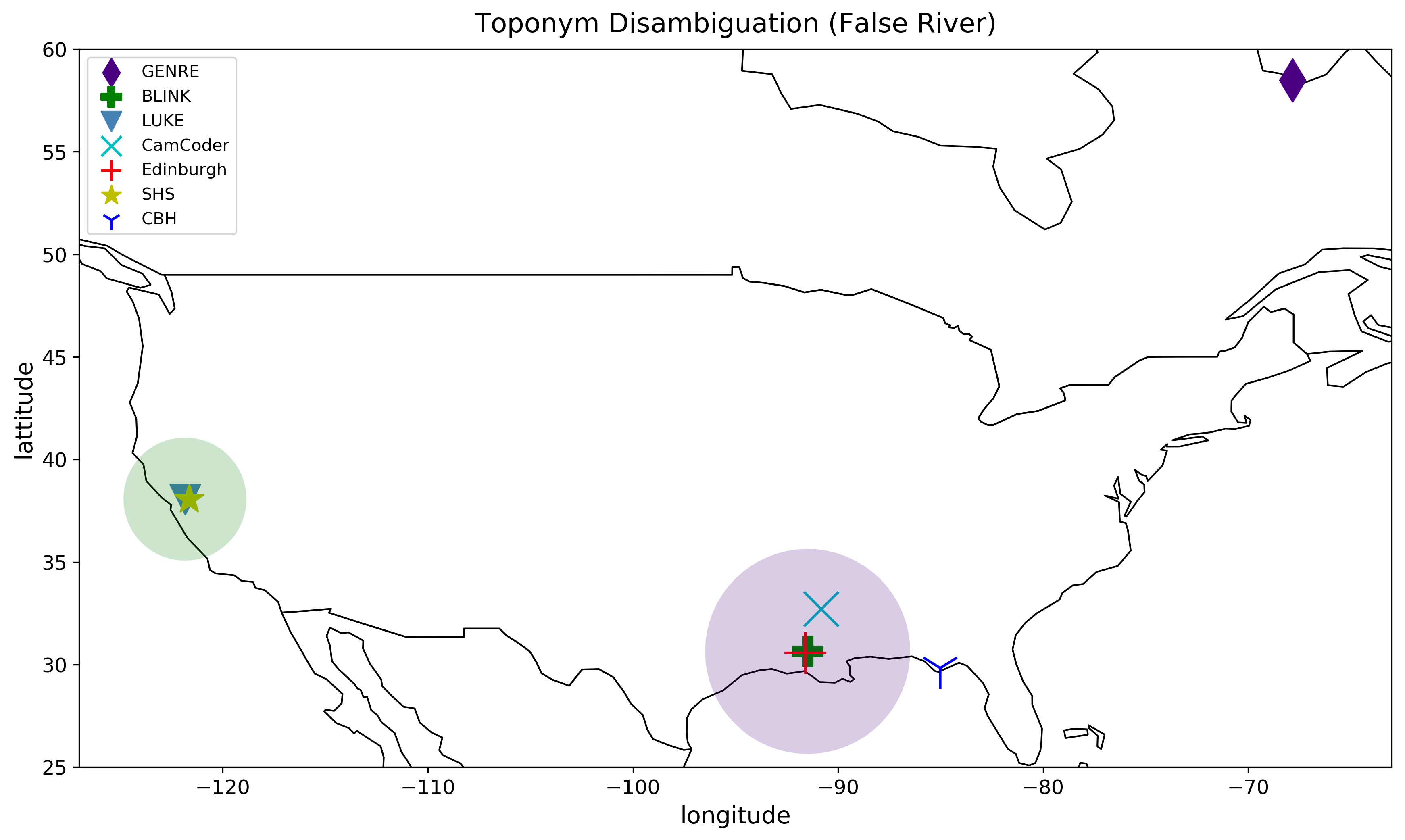}
	  	\caption{An example to show how the voting approach works. The target toponym is \textit{‘False River’}, whose true location is in the largest cluster (purple circle). The context of the toponym is: ‘\textit{The enemy have now left Waterloo, and that is of no importance, but the Rosedale country is of to visit, with the cavalry, and so also is the \textbf{False River} country. The cavalry must go to Rosedale and return by False River}’.}
	  	\label{example2}
	  \end{figure}

 	   	 \begin{figure}[h!] 
	   	 \small
	  	\centering
	  	\includegraphics[width=12cm]{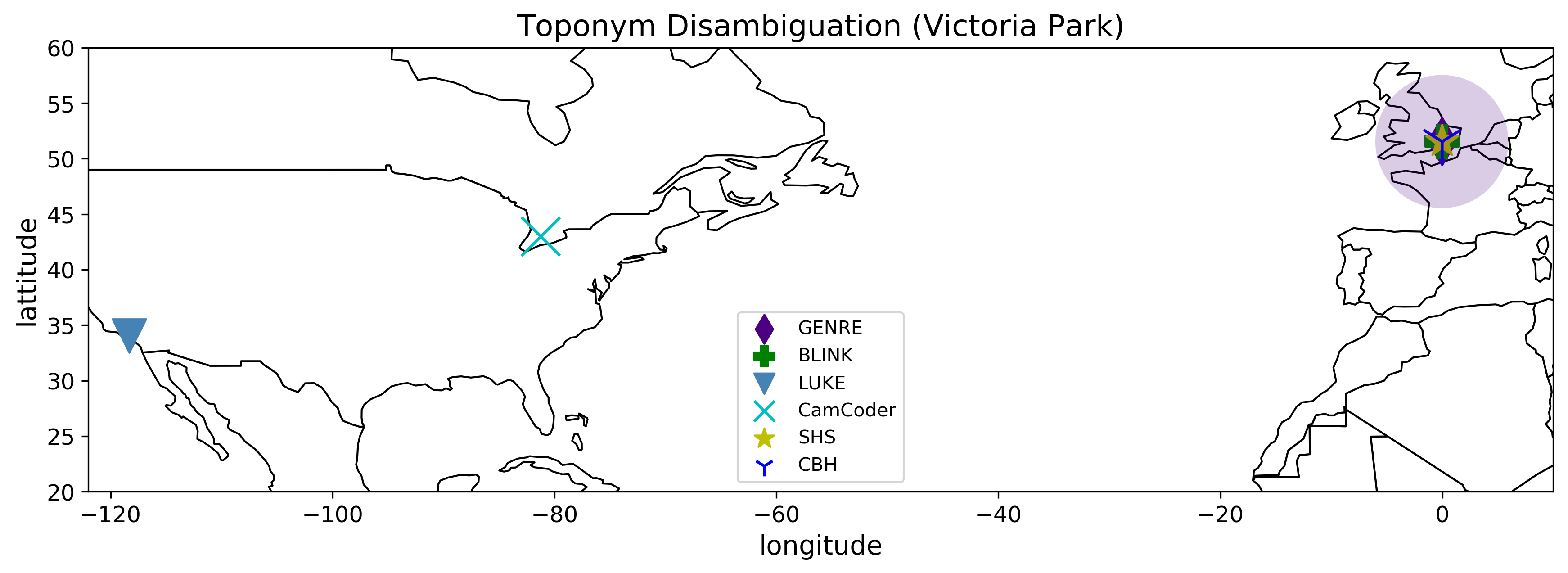}
	  	\caption{An example to show how the voting approach works. The target toponym is \textit{‘Victoria Park’}, whose true location is in the largest cluster (purple circle). The context of the toponym is: ‘\textit{The Clash - White Riot (Live 1978 \textbf{Victoria Park}, London): http://t.co/ylBdgUPwB0 via @YouTube Let's start our shift!}’.}
	  	\label{example3}
	  \end{figure}

 	   	 \begin{figure}[h!] 
	   	 \small
	  	\centering
	  	\includegraphics[width=12cm]{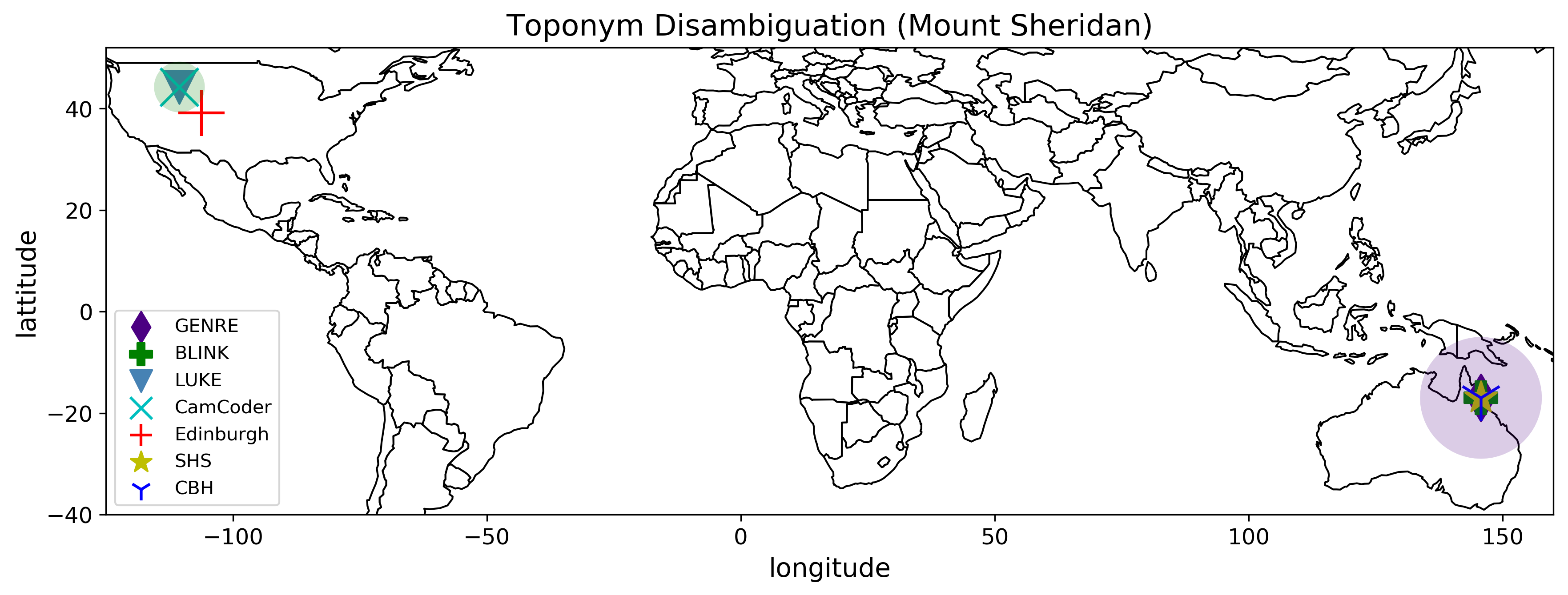}
	  	\caption{An example to show how the voting approach works. The target toponym is \textit{‘Mount Sheridan’}, whose true location is in the largest cluster (purple circle). The context of the toponym is: ‘\textit{Nine cases of the mosquito-borne illness have been confirmed in the Cairns suburbs of Edmonton, \textbf{Mount Sheridan}, Bentley Park, and Trinity Beach.}’.}
	  	\label{example4}
	  \end{figure}


We use four examples to further illustrate the principle of the voting approach. We assume that the voting ensemble combines seven individual approaches with each having one vote: \textbf{GENRE}, \textbf{BLINK}, \textbf{LUKE}, \textbf{CamCoder}, \textbf{Edinburgh Geoparser}, \textbf{SHS}, and \textbf{CBH}. Figure \ref{example1} shows the estimated location of \textit{`SA'} by the seven individual approaches and formed clusters. In GeoNames we can find 58 records of \textit{`SA'}. \textit{`SA'} here refers to the country of South Africa. Its true location is located in the largest cluster, denoted by the purple circle, including the estimations of three approaches. 
 Figure \ref{example2} shows the example of \textit{`False River'}. In GeoNames we can find 23 records of \textit{`False River'}. \textit{`False River'} here refers to a county in Louisiana, US. Its true location is located in the largest cluster, denoted by the purple circle, including the estimation of three approaches. 
  Figure \ref{example3} shows the example of  \textit{`Victoria Park'}. In GeoNames we can find 589 records of \textit{`Victoria Park'}. \textit{`Victoria Park'} here refers to a park in London, UK. Its true location is located in the largest cluster, denoted by the purple circle, including the estimation of four approaches. Edinburgh Geroparser cannot recgonize the toponym. It thus cannot vote. Figure \ref{example4} shows the example of  \textit{`Mount Sheridan'}. In GeoNames we can find 25 records of \textit{`Mount Sheridan'}. \textit{`Mount Sheridan'} here refers to a suburb of Cairns in the Cairns Region, Queensland, Australia. Its true location is located in the largest cluster, denoted by the purple circle, including the estimation of four approaches. 

From the four examples, we can observe that no one individual approach can correctly resolve the toponym in all the examples, while the voting ensemble can correctly resolve all the toponyms.

\section{Experiments}
In this section, we first introduce the used test datasets and evaluation metrics. We then propose a voting ensemble and compare the voting ensemble with the \toolss\space  latest and commonly-used approaches regarding correctness and computational efficiency. Finally, we conduct a sensitivity analysis of the proposed voting approach.

\subsection{Test datasets}

	   	 \begin{figure} [h!] 
	   	 \small
	  	\centering
	  	\includegraphics[width=15cm]{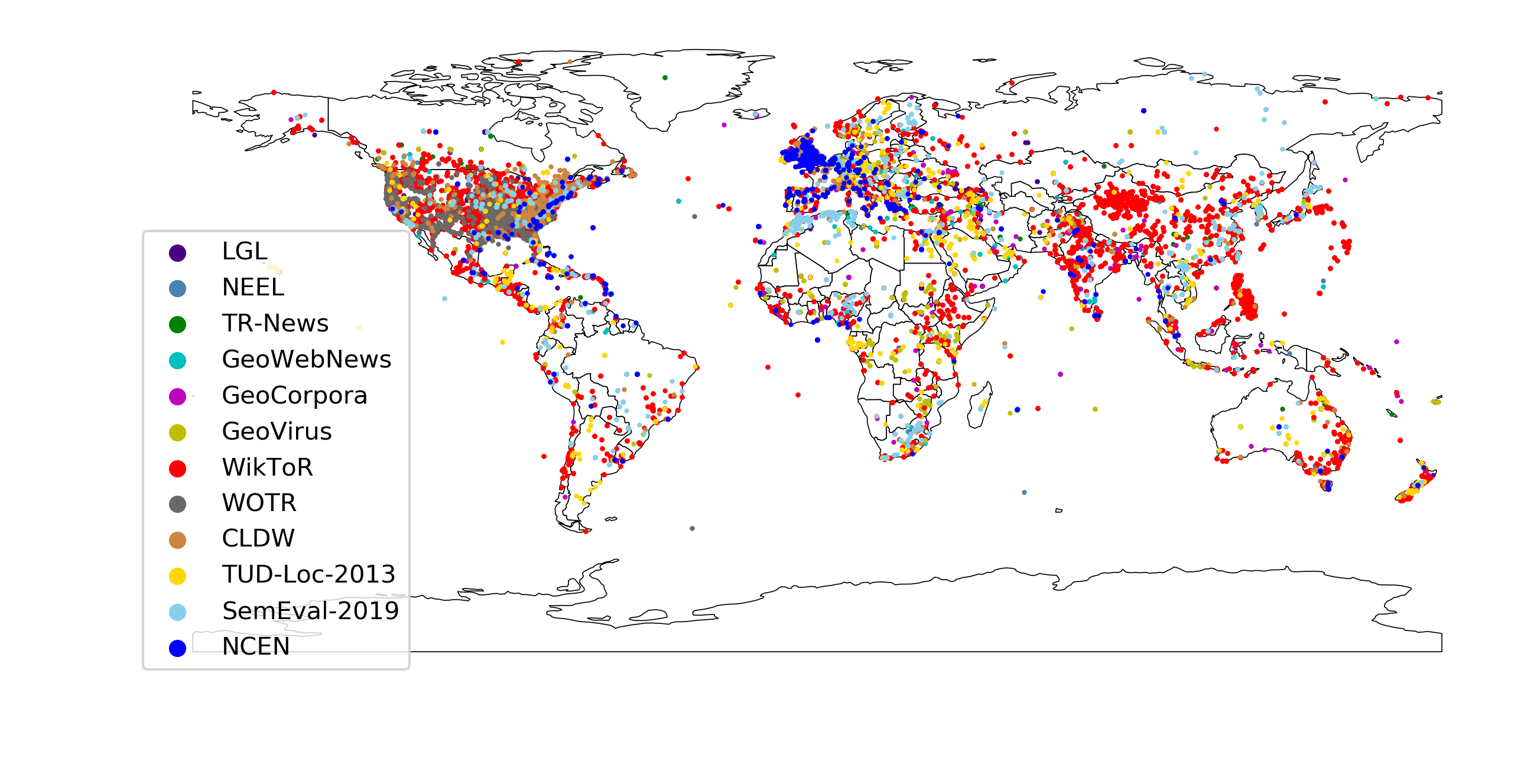}
	  	\caption{Spatial distribution of the \placess\space toponyms in the 12 datasets. }
	  	\label{spatial}
	  \end{figure}

To fairly and thoroughly evaluate the voting ensemble and compare it with the \toolss\space individual approaches, we used \datass\space public datasets, which are of six types, containing \placess\space toponyms across the world. Table \ref{data1} summarizes the characteristics of the test datasets, while Figure \ref{spatial} illustrates the spatial distribution of the toponyms in the datasets. Note that, entity linkers normally use Wikipedia as the target KB, while the toponyms in most of the test datasets are linked to the entries of GeoNames except WikToR, NEEL, GeoVirus, and NCEN. However, the coordinates of some coarse-grained places (e.g., country) in Wikipedia and GeoNames are inconsistent. For instance, \textit{`United States'} is geocoded to (40,-100) and (39.76, -98.5), \textit{`China'} is geocoded to (35, 103) and (35, 105), and \textit{`Russia'} is geocoded to (66, 94) and (60, 100) in Wikipedia and GeoNames, respectively. Such places appear frequently in the datasets, which can cause incorrect evaluation. From the datasets, we recognize 29 frequent and misaligned toponyms, including [\textit{`China', 'Chinese', `Russia', 'Russian', 'Russians', `Australia',  `Canada','Canadians','Canadian', `United States', 'American','USA', 'America', 'U.S.', 'U.S', 'United States of America',' Americans', 'North America', 'South America', `India', `Algeria', `Europe','European', 'Western Europe', `Asia', `Africa', 'West Africa', 'North Africa', 'Middle East'}]. In total, we found 3,147 records from the datasets, and they will be ignored during the evaluation. These toponyms are not ambiguous in the datasets and can be easily resolved by existing approaches.  

\begin{table}[htbp!]
\centering
\footnotesize
\caption{Summary of the test datasets.}
\label{data1}

\begin{tabular}{cccc}
\hline
\textbf{Dataset} & \textbf{Text Count} & \textbf{Toponym Count} & \textbf{Type}       \\ \hline
LGL              & 588                          & 5,088                  & News                \\
NEEL             & 2,135                        & 481                    & Tweet               \\
TR-News          & 118                          & 1,319                  & News             \\
GeoWebNews       & 200                          & 5,121                  & News                \\
GeoCorpora       & 6,648                        & 3,100                  & Tweet               \\
GeoVirus         & 230                          & 2,170                  & News                \\
WikToR           & 5,000                        & 31,500                 & Wikipedia article  \\
WOTR             & 1,643                         & 11,795                 & History             \\
CLDW             & 62                           & 3,814                  & History             \\
TUD-Loc-2013     & 152                          & 3,850                  & Web page           \\
SemEval-2019     & 90                           & 8,360                  & Scientific article \\
NCEN             & 455                          & 3,364                  & History             \\ \hline
\end{tabular}
\end{table}




Details of the \datass\space test datasets are as follows: 
\begin{itemize}

\item \textbf{LGL \footnote{\url{https://github.com/milangritta/Pragmatic-Guide-to-Geoparsing-Evaluation/blob/master/data/Corpora/lgl.xml}}:} Local-Global Lexicon (LGL) corpus was created by \citet{lieberman2010geotagging}. Toponyms were manually annotated and linked to entries in GeoNames from 588 human-annotated news articles published by 78 local newspapers.

     \item \textbf{NEEL \footnote{\url{http://microposts2016.seas.upenn.edu/challenge.html}}}: It is the gold dataset of 2016 Named Entity rEcognition and Linking (NEEL) challenge. The dataset includes event-annotated tweets covering multiple noteworthy events from 2011 to 2013, such as the death of Amy Winehouse, the London Riots, and the Westgate Shopping Mall shootout. Entities of different types, such as \textit{Location}, \textit{Person}, \textit{Organization}, \textit{Event}, and \textit{Product} were  annotated and linked to entries in DBpedia. 

\item \textbf{TR-News \footnote{\url{https://github.com/milangritta/Pragmatic-Guide-to-Geoparsing-Evaluation/blob/master/data/Corpora/TR-News.xml}}:} TR-News was created by \citet{kamalloo2018coherent}. Toponyms were manually annotated and linked to entries in GeoNames from  118 news articles of various news sources.

\item \textbf{GeoWebNews \footnote{\url{https://github.com/milangritta/Pragmatic-Guide-to-Geoparsing-Evaluation/tree/master/data}}:} The dataset was shared by \citet{gritta2018melbourne}, which comprises 200 human-annotated news articles from 200 globally distributed news sites collected from April 1st to 8th in 2018. The toponyms were linked to  entries in GeoNames.

      \item \textbf{GeoCorpora \footnote{\url{https://github.com/geovista/GeoCorpora}}:} It was created by \citet{wallgrun2018geocorpora}. Toponyms in tweets were  annotated and linked to the entries of GeoNames. The dataset corresponds to multiple noteworthy events (e.g., earthquake, ebola, fire, flood, and rebel) that happened across the world in 2014 and 2015. 
      
\item \textbf{GeoVirus \footnote{\url{https://github.com/milangritta/Pragmatic-Guide-to-Geoparsing-Evaluation/blob/master/data/Corpora/GeoVirus.xml}}:}
GeoVirus was created by \citet{gritta2018melbourne}. Toponyms were manually annotated and linked to Wikipedia from 239 news articles related to disease outbreaks and epidemics, such as Ebola, Bird Flu, and Swine Flu. 

\item \textbf{WikToR \footnote{\url{https://github.com/milangritta/Pragmatic-Guide-to-Geoparsing-Evaluation/blob/master/data/Corpora/WikToR.xml}}:} The dataset was created by \citet{gritta2018s} in an automatic manner. In WikToR, a text corresponds to a geographically annotated Wikipedia article entitled with a toponym. Only the toponyms in the text, which are the same as the title are automatically annotated. WikToR contains 5,000 articles with many ambiguous toponyms, such as (\textit{Santa Maria, California}), (\textit{Santa Maria, Bulacan}), (\textit{Santa Maria, Ilocos Sur}), and (\textit{Santa Maria, Romblon}).

\item \textbf{WOTR \footnote{\url{https://github.com/barbarainacioc/toponym-resolution/tree/master/corpora/WOTR}}:} The dataset was created by \citet{delozier2016creating} based on a set of American Civil War archives, known as \textit{Offical Records of the War of the Rebellion}. WOTR contains two levels of geolocation: document-level and toponym-level. In this study, we only use the toponym-level annotations.

\item \textbf{CLDW \footnote{\url{https://github.com/UCREL/LakeDistrictCorpus}}:} The Corpus of Lake District Writing (CLDW) was created by \citet{rayson2017deeply} based on 80 texts which are writing samples about the English Lake District between the early seventeenth and the beginning of the twentieth century.

\item \textbf{TUD-Loc-2013 \footnote{\url{https://bitbucket.org/palladian_pk/tud-loc-2013/src/master/}}:} The dataset  was first utilized in \citep{katz2013learn}, consisting of 152 texts obtained from  web pages. 

    \item \textbf{SemEval-2019-12 \footnote{\url{https://github.com/TharinduDR/SemEval-2019-Task-12-Toponym-Resolution-in-Scientific-Papers}}:} It is the gold dataset of the Task 12 (Toponym Resolution in Scientific Papers) of the 13th International Workshop on Semantic Evaluation (SemEval) \citep{weissenbacher2019semeval}. Full-text journal articles were downloaded from the subset of PubMed Central (PMC) and toponyms in the articles were manually annotated and linked to entries in GeoNames.

\item \textbf{NCEN \footnote{\url{https://bl.iro.bl.uk/concern/datasets/f3686eb9-4227-45cb-9acb-0453d35e6a03}}:} The Nineteenth-Century English Newspapers (NCEN) dataset  was created by \citet{ardanuy2022dataset}, containing 343 annotated articles from newspapers based in four different locations in England (i.e., Manchester, Ashton-under-Lyne, Poole, and Dorchester). The articles were published between 1780 and 1870. 3,364 toponyms were annotated, of which 2,784 were linked to Wikipedia.

\end{itemize}

\subsection{Evaluation metrics}

To fairly evaluate toponym resolution approaches, we assume that all the toponyms in the datasets can be correctly recognized at the toponym recognition step. 
However, DBpedia Spotlight and Edinburgh Geoparser provide only an online API and deploy the toponym recognition module on servers. Therefore, when evaluating the correctness of the DBpedia Spotlight and Edinburgh Geoparser, we will compare them with the other approaches on the correctly recognized toponyms (a subset of gold toponyms) by the two approaches, respectively.

From the standard metrics defined in \citep{gritta2020pragmatic} for evaluating the toponym disambiguation approaches, we adopt the three most important metrics. They are: (1) \textit{Accuracy@161km}, which is the percentage of geocoding errors that are smaller than 100 miles (161 km); (2) \textit{Mean Error (ME)}, which is the sum of the distance error of all the toponyms in a dataset divided by the number of toponyms in the dataset;  (3) \textit{Area Under the Curve (AUC)}, which is the total area under the curve of the normalized log error distance. AUC is calculated using Equation \ref{auc_for}, where $x_i$ denotes the distance error of the $i$-th toponym, $N$ denotes the count of toponyms, and 20039 is the maximum possible error in km on earth. 
\begin{equation}
\label{auc_for}
    AUC = \int_{i=1}^{N}\frac{ln(x_i+1)}{ln(20039)}dx
\end{equation}

\subsection{Voting ensemble}
 Given \toolss\space individual approaches, there are many possible combination manners. We propose an optimal voting ensemble that combines seven individual approaches and manually assign a weight (count of votes) to each approach, denoted by the number in the brackets. We decide the combination manner and weights according to numerous experimental results. For the voting algorithm, we set the DBSCAN parameters $minPts$ and $eps$ to 2 and 10 (km), respectively. In the section on sensitivity analysis, we will investigate the importance of an individual approach in voting ensembles and the impact of different parameter settings on disambiguation performance. 
 The voting ensemble is as follows:
 \begin{itemize}
     \item \textbf{Voting}: GENRE (3), BLINK (2), LUKE (2), CamCoder (1), SHS (1), CBH (1), Edinburgh Geoparser (1).
      
 \end{itemize}



	   	 \subsection{Results}
\begin{figure}[h!]
	   	 \small
	  	\centering
	  	\includegraphics[width=14cm]{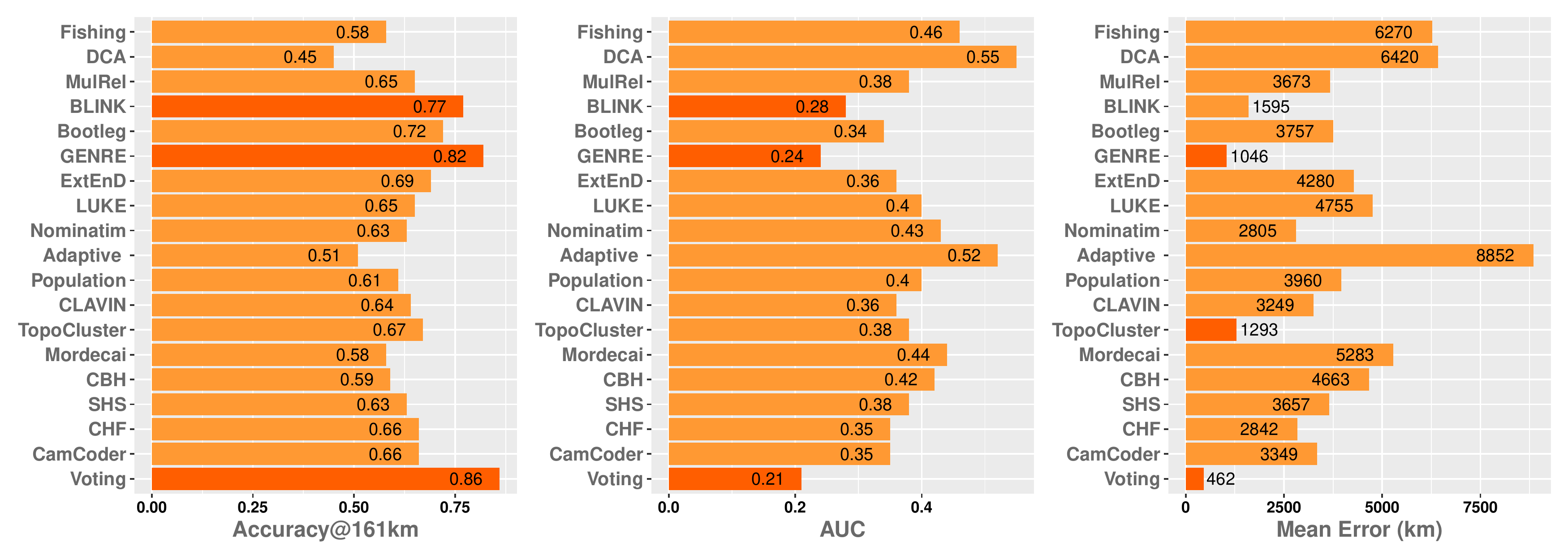}
	  	\caption{Average \textit{Accuracy@161km} ($\uparrow$), \textit{AUC}  ($\downarrow$),  and  \textit{ME} ($\downarrow$) of  approaches on gold toponyms. The three best  approaches on each metric are highlighted.}
	  	\label{result1}
	  \end{figure}

	   	 Tables \ref{acc161all}, \ref{aucall}, \ref{meall} in the Appendix present the results of the 18 individual approaches (excluding Edinburgh Geoparser and DBpedia Spotlight) and the voting ensemble on the gold toponyms of each test dataset.
	   	 We can see that the voting ensemble achieves the best result on 35/36 (3 metrics evaluatetd for \datass\space datasets) indicators. We then average each metric of the approaches on the \datass\space datasets. The result is shown in Figure \ref{result1}. We can observe that the voting ensemble achieves an \textit{Accuracy@161km} at 0.86, improving the best individual approach, GENRE, by 5\%. Similarly, the  ensemble achieves the best  \textit{ME} and \textit{AUC}, improving 
	   	the best individual approach,  GENRE, by 57\% and 13\%, respectively. 
	   	 
\begin{figure}[h!]
	   	 \small
	  	\centering
	  	\includegraphics[width=14cm]{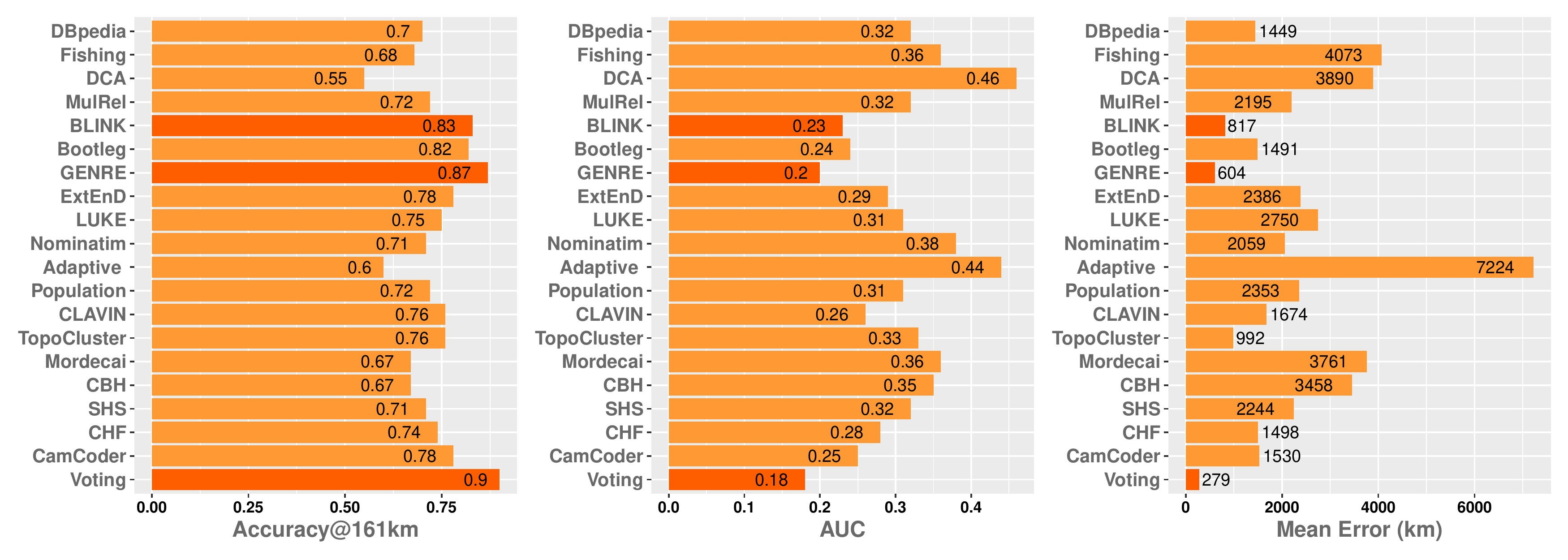}
	  	\caption{Average \textit{Accuracy@161km} ($\uparrow$), \textit{AUC}  ($\downarrow$),  and  \textit{ME} ($\downarrow$) of the approaches  on the subset of gold toponyms, which are correctly recognized by DBpedia Spotlight. The  top three approaches on each metric are highlighted.}
	  	\label{result2}
	  \end{figure}

\begin{figure}[h!]
	   	 \small
	  	\centering
	  	\includegraphics[width=14cm]{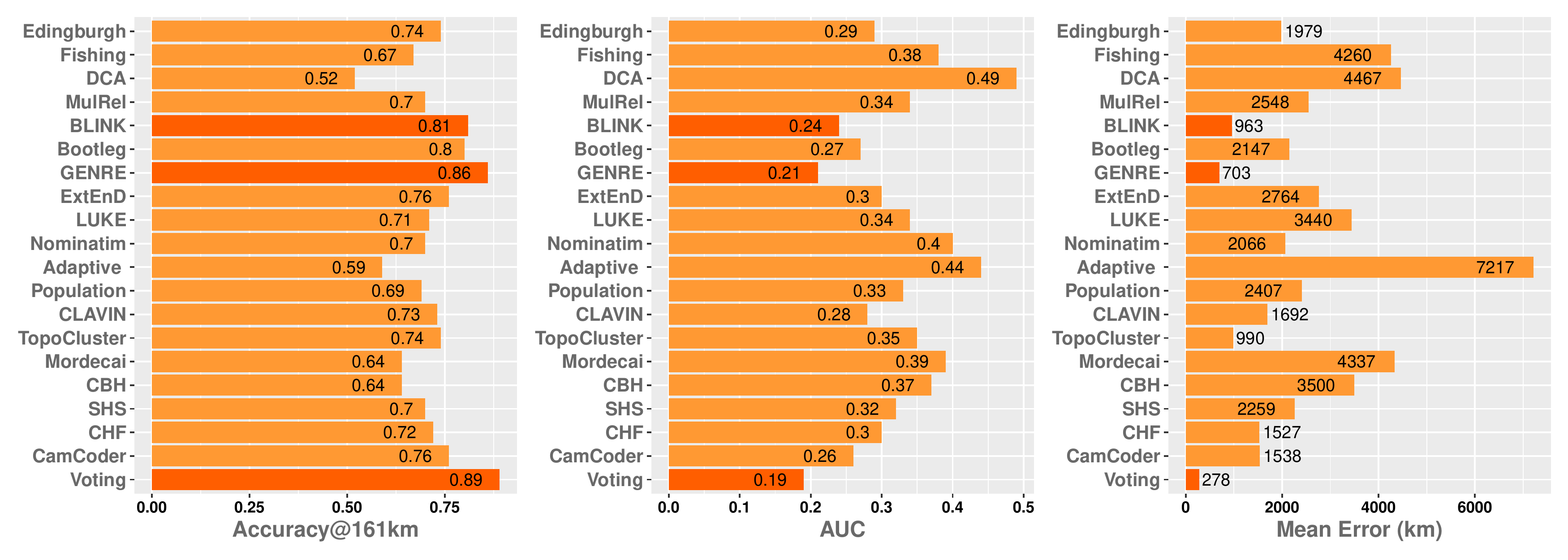}
	  	\caption{Average \textit{Accuracy@161km} ($\uparrow$), \textit{AUC}  ($\downarrow$),  and  \textit{ME} ($\downarrow$) of the approaches on the subset of gold toponyms, which are correctly recognized by Edinburgh Geoparser. The top three approaches on each metric are highlighted.}
	  	\label{result3}
	  \end{figure}

\begin{figure}[h!]
	   	 \small
	  	\centering
	  	\includegraphics[width=12cm]{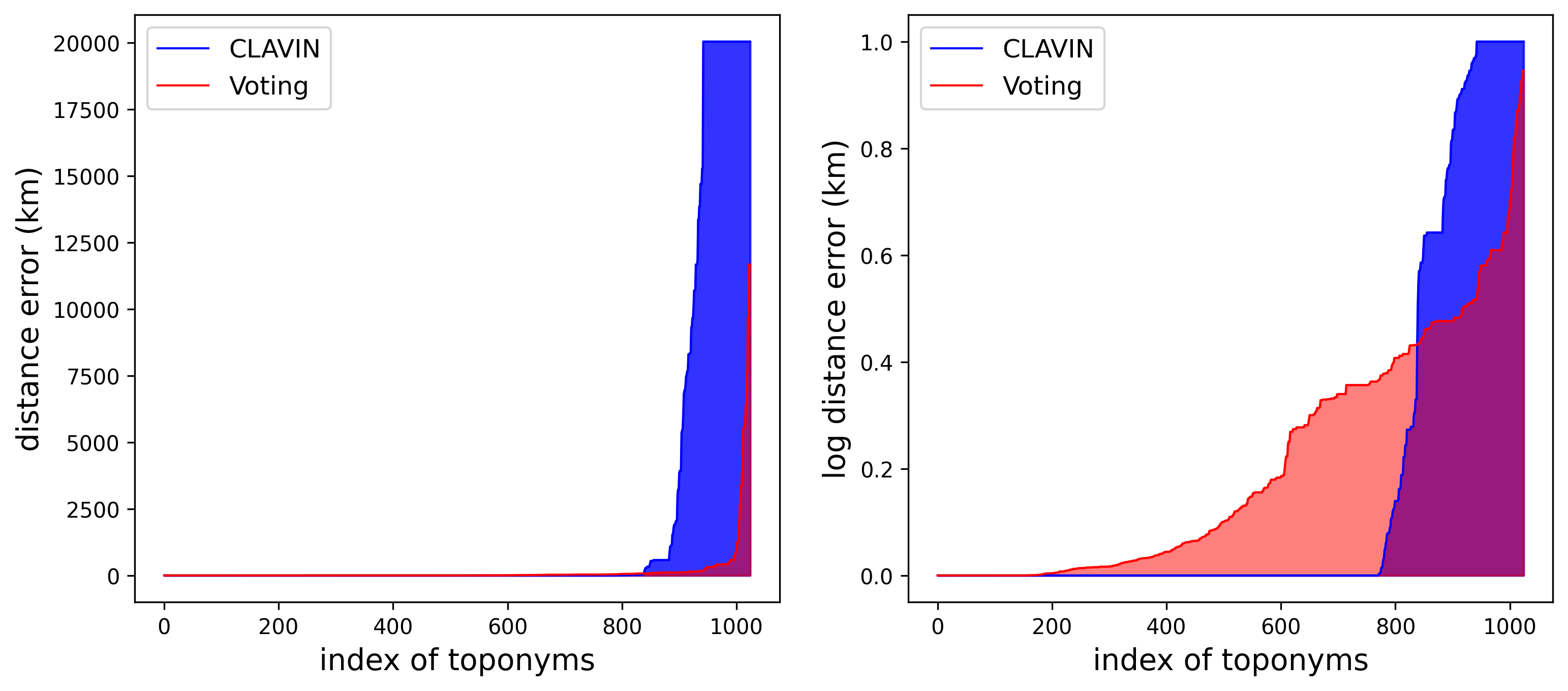}
	  	\caption{AUC of CLAVIN and the Voting ensemble on the subset of gold toponyms of GeoWebNews (correctly recognized by DBpedia Spotlight)  before and after using the natural logarithm.}
	  	\label{auc}
	  \end{figure}

	   	 Tables \ref{acc161db},  \ref{aucdb}, and \ref{medb} in the Appendix present the raw results of the 19 individual approaches and the voting ensemble on the subset (51\%) of gold toponyms, which are correctly recognized by DBpedia Spotlight. This is to evaluate DBpedia Spotlight and compare it with other approaches based on the same set of toponyms.  
	   	 We can see that the voting ensemble achieves the best result on 26/36 indicators. Figure \ref{result2} shows the average \textit{Accuracy@161km}, \textit{Mean Error}, and \textit{AUC}, while the voting ensemble  performs the best. Tables \ref{acced}, \ref{auced}, and \ref{meed}  present the raw results of the individual approaches and the voting ensemble on the subset (59\%)  of gold toponyms, which are correctly recognized by Edinburgh Geoparser. This is to evaluate Edinburgh Geoparser and compare it with other approaches based on the same set of toponyms. 
	   	The voting ensemble achieves the best result on 27/36 indicators. Figure \ref{result3} shows the average \textit{Accuracy@161km}, \textit{Mean Error}, and \textit{AUC}, and the voting ensemble still performs the best. In a nutshell, the voting ensemble still performs the best compared with individual approaches on the subset of the gold toponyms. However, the achieved result is not as good as on the gold toponyms, i.e., 26/36 and 27/36 Versus 35/36. 
	   	 This is mainly because the coordinates of a place in GeoNames and Wikipedia are inconsistent, while the toponyms in most of the datasets are linked to GeoNames. Although a toponym is correctly resolved by entity linkers, such as GENRE, the distance error can be still high. For example, the distance between the coordinates of France (the country in Europe) in Wikipedia (47,2) and in GeoNames (36,2) is 111 km. When the approaches are evaluated on only half of the gold toponyms, the  performance is drastically affected by the inconsistent coordinate issue. This especially affects the result of AUC, in which, the distance error is scaled down using the natural logarithm, which increases the weight or importance of small distance errors in calculating the AUC. This results in the higher AUC of entity linkers and the voting ensemble on less ambiguous datasets (i.e., GeoWebNews, GeoCorpora, and SemEval) than that of toponym resolution approaches, such as CLAVIN. We can observe from Figure \ref{auc} that before using natural logarithm, the AUC of the voting ensemble is far less than that of CLAVIN, while the result reverses after using the natural logarithm. When the original distance error is 70 km, the log distance error is already 0.42, nearly half of the log distance error of the original distance error at 7000 km.  
	   	 
	   	  From Figures \ref{result1}, \ref{result2}, and \ref{result3} we can observe that, among individual approaches, the state-of-the-art entity linkers, GENRE and BLINK, achieve promising results, outperforming the other entity linkers as well as toponym resolution approaches, although the former also disambiguate other types of entities rather than just toponyms. From the tables in Appendix, we can see that GENRE and BLINK are especially effective on highly ambiguous datasets, such as WikToR, WOTR, and LDC, performing much better than the other individual approaches. These datasets contain many less-common or low-frequency places, such as \textit{`Paris, Missouri'} and \textit{`Lima, Oklahoma'}. The two baseline systems, Nominatim and Population-Heuristics, which adopt simple heuristics (i.e., popularity), thus perform  poorly on these datasets. 
	   	  Regardless of whether highly ambiguous datasets or general datasets, the voting ensemble that combines several individual approaches can always achieve state-of-the-art performance, proving its generalizability and robustness in toponym disambiguation.


\subsection{Place category}


In this experiment, we investigate the disambiguation performance of the approaches on different types of places. We divide places into four categories: admin units (e.g., continent, country, state, and county), POIs (e.g., park, church, school, and hospital), traffic ways (e.g., street, road, highway, and bridge), natural features (e.g., river, creek, beach, and hill). 
In the datasets of GeoCorpora, LGL, TR-News, GeoWebNews, CLDW, and Semeval-2019, the GeoNames ID of places have been provided, through which we can determine the category of the places. We treat the places whose feature codes \footnote{\url{http://www.geonames.org/export/codes.html}} are \textit{A} (e.g., country, state, and region) and \textit{P} (e.g., city and village) in GeoNames as admin units, the places whose feature codes are \textit{L} (e.g., park and port) and \textit{S} (e.g., sport, building, and farm) as POIs,  the places whose feature codes are \textit{H} (e.g., stream and lake), \textit{T} (e.g., mountain, hill, and rock), \textit{U} (e.g., undersea and valley), and \textit{V} (e.g., forest and grove) as natural features, and the places whose feature codes are \textit{R} (e.g., road and railroad) as traffic ways.
Moreover, to increase the number of POIs, natural features, and traffic ways, we rule that the toponyms that contain the words of [\textit{  `church',  `hospital',  `school', `university',  `park'}] are potential POIs, that contain the words of  [\textit{`river', `creek'}] are potential natural features, and that contain the words of [\textit{`street', `road', `roads', `railroad', `highway', `way', `drive', `hwy', `bridge', `trail'}] are potential traffic ways. We searched all the datasets for the three types of places. We then manually checked the annotation and removed incorrect ones. Finally, we determined in total 13,878 admin units (i.e., \textit{`EU'}, \textit{`Berlin'}, and \textit{`Boone County'}), 820 POIs (e.g., {\textit{`Lambert-St. Louis International Airport'}, \textit{`Sam Houston High School'}, and \textit{`westboro baptist church'}}), 1,605 natural features (\textit{`Pine Island Bayou', `Skiddaw Mountain'}, and \textit{`Little Pine Creek'}), and 336 traffic ways (i.e., \textit{`High Street', `Lynchburg Railroad bridge'}, and \textit{`Highway 49'}).

\begin{figure}[h!]
	   	 \small
	  	\centering
	  	\includegraphics[width=11cm]{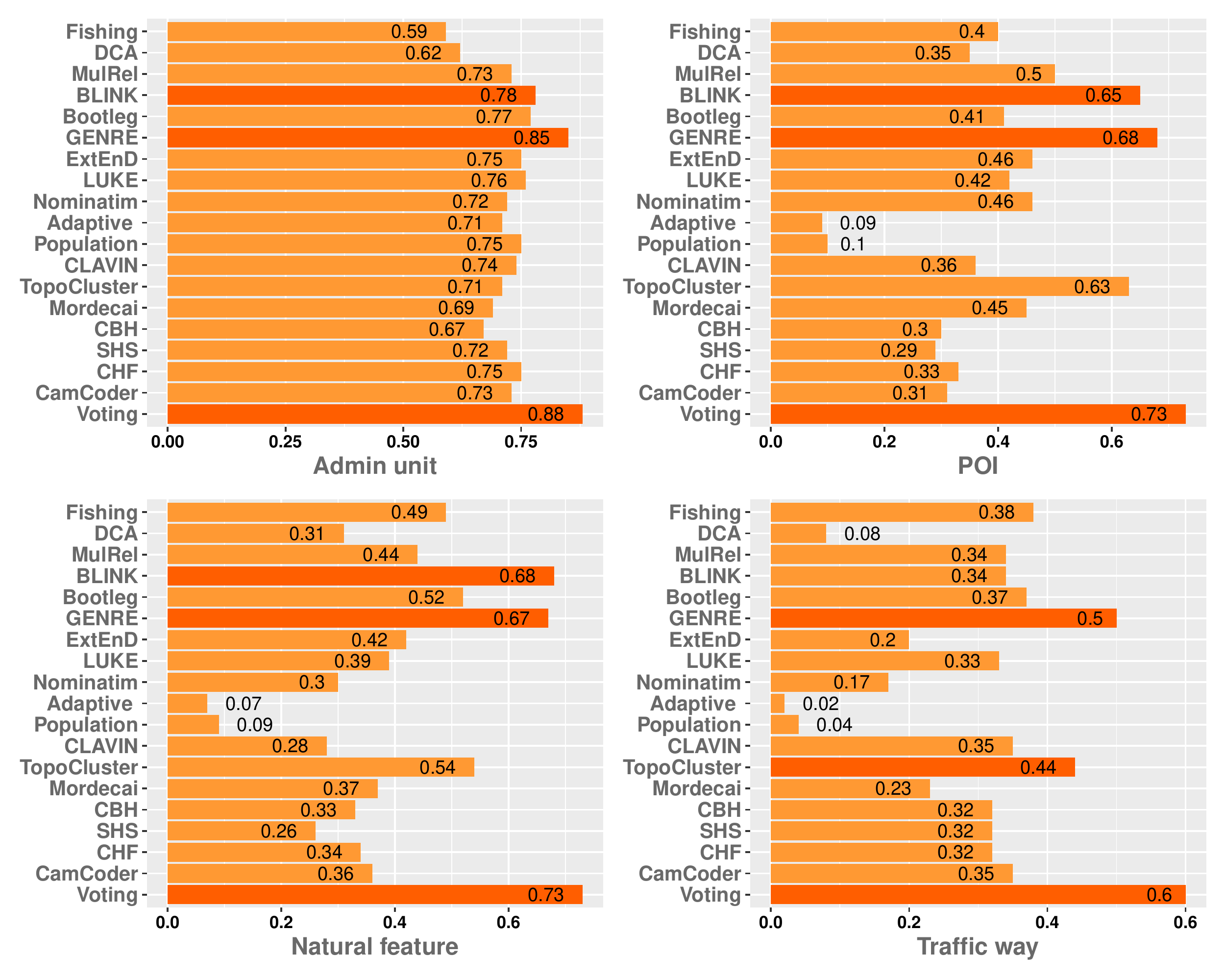}
	  	\caption{  \textit{Accuracy@161km} of the approaches on four categories with 13,878 admin units (i.e., \textit{`EU'}, \textit{`Berlin'}, and \textit{`Boone County'}), 820 POIs (e.g., {\textit{`Lambert-St. Louis International Airport'}, \textit{`Sam Houston High School'}, and \textit{`westboro baptist church'}}), 1605 natural features (\textit{`Pine Island Bayou', `Skiddaw Mountain'}, and \textit{`Little Pine Creek'}), and 336 traffic ways (i.e., \textit{`High Street', `Lynchburg Railroad bridge'}, and \textit{`Highway 49'})}
	  	\label{category}
	  \end{figure}

We then calculated \textit{ Accuracy@161km} of the approaches on each type of places. Note that, we rule out Edinburgh Geoparser and DBpedia Spotlight since we cannot import gold toponyms to the approaches and their own toponym recognition modules can only correctly recognize a small proportion of places, especially POIs, traffic ways, and natural features, which can cause unfair evaluation. For example, Edinburgh Geoparser can only recognize 21 traffic ways and DBpedia Spotlight can only recognize 13 traffic ways from all 336 traffic ways. 
We can observe from Figure \ref{category} that most of the approaches show superior performance on resolving coarse-grained places (i.e., admin units), with 15 of 19 can correctly resolving over 70\% of the admin units. However, most of them are incapable of resolving fine-grained places. Only 4 of 19, 3 of 19, and 1 of 19 can correctly resolving over 60\% of the POIs, natural features, and traffic ways, respectively. The last three categories refer to much more local geographic scopes than admin units and are thus valuable in many key applications, such as emergency rescue and traffic event detection. 
The voting ensemble performs the best, which can correctly resolve over 60\% of the toponyms in all four categories. On average, the voting ensemble improves the best individual approach, GENRE, by 11\% in resolving fine-grained places. However, there is still space for improving the performance of resolving fine-grained places. 

\subsection{Computational efficiency}
In this section, we further investigate the computational efficiency (i.e., speed) of different approaches. In many applications, the texts that need to be geoparsed are of huge volumes, such as some major historical books and reports (e.g., the Old Bailey Online) that each comprises many millions or even billions of words \citep{gregory2015geoparsing} and over millions of tweets related to a crisis event \citep{qazi2020geocov19}. This requires a rapid geoparsing procedure and the speed is thus a critical indicator. 

\begin{figure}[htbp!]
	   	 \small
	  	\centering
	  	\includegraphics[width=12cm]{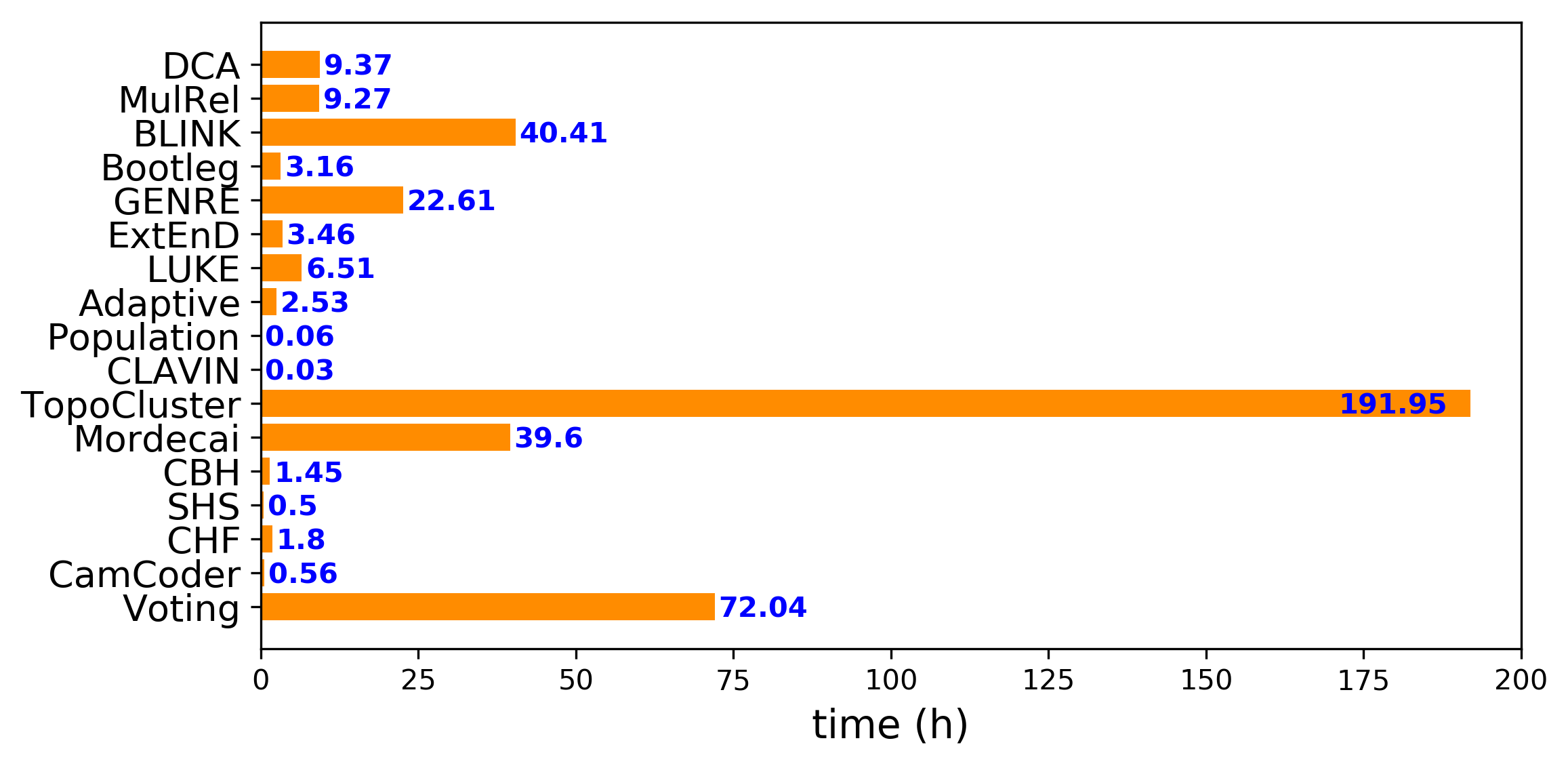}
	  	\caption{Time consumption of the approaches running on the total test datasets. }
	  	\label{computation}
	  \end{figure}

We run each approach on the total datasets and record the consumed time. We do not count the consumed time during the training phase if an approach needs to be trained since it can be conducted offline and is a one-time process. Note that, we do not include Edinburgh Geoparser, Nominatim, DBpedia Spotlight, and Entity-Fishing in the comparison since they are online services and it is impossible to count the amount of time of processing done on the server.
 We run the toponym resolution approaches on a Dell laptop with an Intel Core i7-8650U CPU (1.90 GHz 8-Core) and a RAM of 16 GB, while we run the entity linkers on an NVIDIA Tesla V100 GPU of a cluster node since they require a GPU execution environment. Figure \ref{computation} illustrates the consumed time of different approaches required to process all datasets.

We can observe that the speed of different approaches varies drastically. It takes 2 minutes to 191 hours for these approaches to process the total datasets. Generally, the entity linkers take much more time (from 3 hours to 40 hours) than toponym resolution approaches (from 2 minutes to 2.5 hours) except TopoCluster and Mordecai since the former was normally built on large language model, such as BERT and deals with more complex issues (disambiguating not only toponyms but also other types of entities) than the latter. 
Regarding individual approaches, TopoCluster is the slowest, taking nearly 191 hours while CLAVIN is the fastest, taking only 2 minutes. The time consumption for a voting ensemble equals the sum of the time of every individual approach that it combines. Therefore, the voting ensemble takes 72 hours for resolving \placess\space toponyms,  which means on average resolving a toponym takes 2.6 seconds. There is a trade-off between correctness and speed.

	   	 \subsection{Sensitivity analysis}
	   	 In this section, we investigate how the removal of an individual approach and different parameter settings of the voting approach can affect the disambiguation performance of voting ensembles.

	   	 	   	 \subsubsection{Configuration}
\begin{figure}[htbp!]
	   	 \small
	  	\centering
	  	\includegraphics[width=15cm]{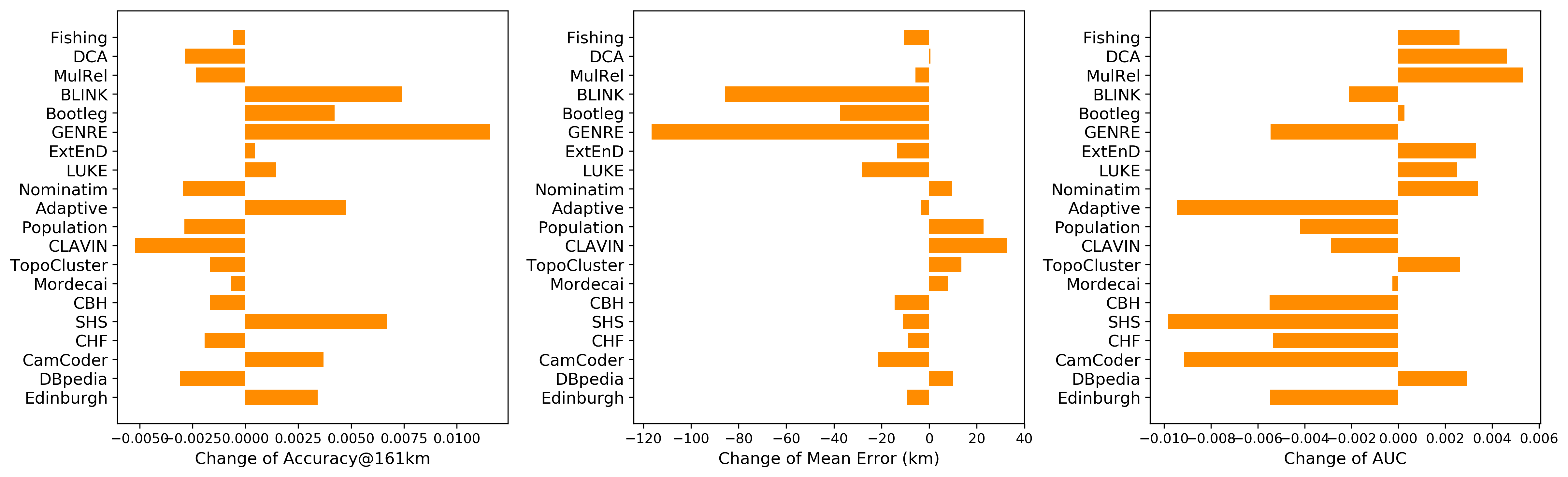}
	  	\caption{Change of \textit{Accuracy@161km}, \textit{AUC}, and \textit{ME} when adding an approach to a voting ensemble. }
	  	\label{importance}
	  \end{figure}

	   	We first investigate how the removal of an individual approach would affect the performance of  voting ensembles. To realize this, a basic voting ensemble is first proposed, which includes all the  \toolss\space individual approaches with each approach having one vote. To obtain the contribution of an individual approach, a degraded ensemble is then constructed by removing the approach from the basic ensemble. We then subtract the  average \textit{ME, Accuracy@161km}, and \textit{AUC} achieved by the degraded ensemble from that of the basic ensemble. The result of all the individual approaches is shown in Figure \ref{importance}. We can see that, regarding \textit{Accuracy@161km}, GENRE, BLINK, and SHS make the largest positive contribution, while CLAVIN and DBpedia Spotlight make the largest negative contribution. Regarding \textit{ME}, GENRE and BLINK make the largest positive contribution, while CLAVIN and Population-Heuristics make the largest negative contribution. 
Regarding \textit{AUC}, CamCoder, SHS, and Adaptive Learning make the largest positive contribution, while DCA and MulRel make the largest negative contribution. Generally, the disambiguation ability of each single approach determines their contribution to the voting ensemble, such as GENRE (with high disambiguation ability) and Population-Heuristics (with low disambiguation ability), which contribute positively and negatively, respectively. 

	   	 \subsubsection{Parameters}
	   	 	 We then analyze the impact of different parameter settings of the voting algorithm on the performance of the voting ensemble. We calculate the average \textit{ME, Accuracy@161km}, and \textit{AUC} of the voting ensemble on the gold toponyms of all the datasets as the change of the parameters. 	   	  The experimental results in this section help us set the optimal parameters of the voting approach. 

	   	 \begin{figure}[h!]
	   	 \small
	  	\centering
	  	\includegraphics[width=12cm]{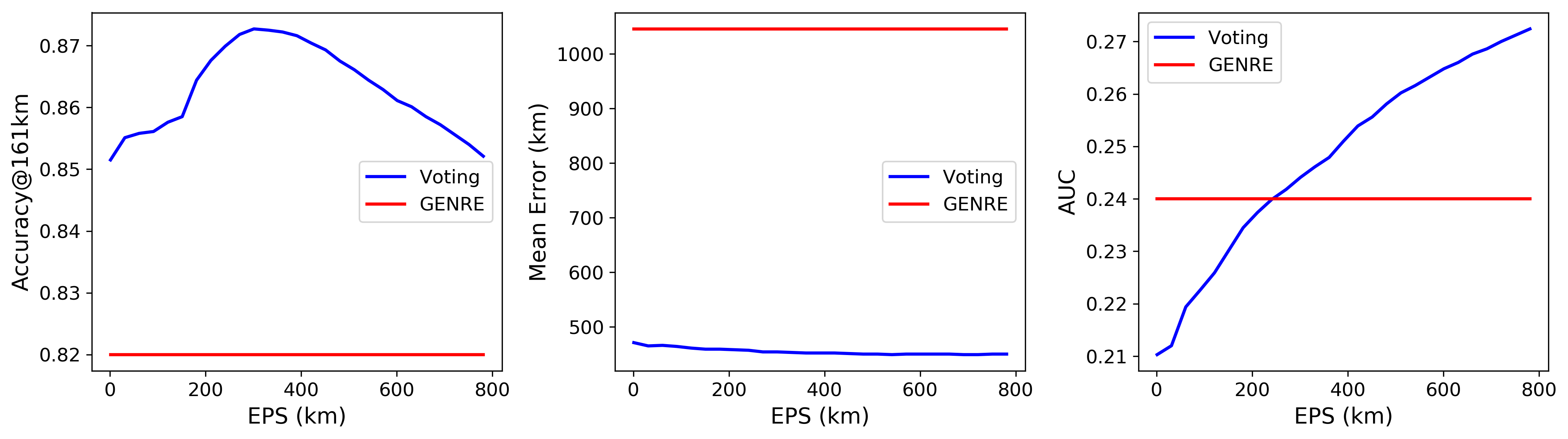}
	  	\caption{Impact of $eps$ on the performance of the voting ensemble. }
	  	\label{eps}
	  \end{figure}
	  	   	 	  In the first experiment, the DBSCAN parameter $eps$ was defined as $eps \in {1, ... , 800 }$ and a step size of 30. Figure \ref{eps} shows the performance of the voting ensemble as the change of $eps$. The red line denotes the performance of the best individual approach, GENRE. We can see that $eps$ has distinct impact on \textit{ME, Accuracy@161km}, and \textit{AUC}. The best \textit{Accuracy@161km} is achieved when $eps$ is set to 350 km, while as the increase of $eps$, \textit{ME} decreases slightly from 470 km to 450 km, and \textit{AUC} increases rapidly from 0.21 to 0.27.
	  	   	 	  
\begin{figure}[h!]
	   	 \small
	  	\centering
	  	\includegraphics[width=12cm]{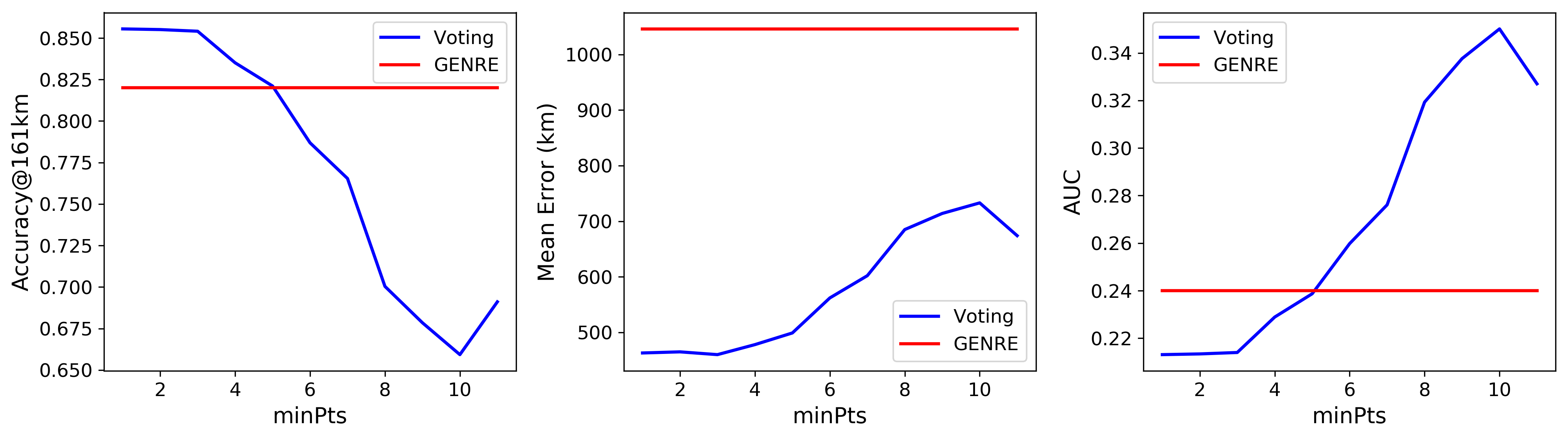}
	  	\caption{Impact of $minPts$ on the performance of the voting ensemble. }
	  	\label{$minPts$}
	  \end{figure}

	  	   	 	  In the second experiment, the DBSCAN parameter $minPts$ was defined as $minPts \in {1, ... , 11 }$ and a step size of 1. Figure \ref{$minPts$} shows the result of the voting ensemble as the change of $minPts$. We can see $minPts$ has a large impact on the performance of the voting ensemble. When $minPts$ is set to 1 and 2, the best performance is reached. The higher the $minPts$, the worse the voting ensemble performs. 


\section{Conclusion}
      In this paper, we investigate if and how voting ensembles that combine several individual approaches can push state-of-the-art performance of toponym disambiguation. Experimental results on \datass\space public datasets of six types (namely, tweets, historical documents, news, web pages, scientific articles, and Wikipedia articles) prove the generalizability and robustness of the voting approach. The deep learning-based entity linkers (i.e., GENRE and BLINK) that are pretrained on nearly 10 million Wikipedia entities show impressive disambiguation performance, performing much better than toponym resolutions approaches. However, there is a trade-off between correctness and speed since the voting approach and the two entity linkers take much more time than most of the other approaches. On average, it takes 2.6 seconds for the voting approach to resolve a toponym. Moreover, there is still space for improving the performance of resolving fine-grained places, such as POIs, natural features, and traffic ways. This will be one of our future research tasks. Furthermore, the idea of voting approaches can be extended to the more general issue, such as entity linking since combining several entity linkers can improve the disambiguation performance for toponyms, and might also for other entities, such as person, and organization.
      


      \section{Data and code availability}
	The code and data that support the findings of this study is available in GitHub with the link \href{https://github.com/uhuohuy/toponym-disambiguation-voting}{https://github.com/uhuohuy/toponym-disambiguation-voting}.  

\section{Competing interests}
The authors declare no competing interests.

	\bibliographystyle{tfcad}
	\bibliography{placeName}
	
	\appendix
\section{} \label{appendix:raw}

	\begin{table}[h!]
		\caption{\textit{Accuracy@161km} on the gold toponyms of all the datasets. TN denotes TR-News,  GWN denotes GeoWebNews, GC denotes GeoCorpora, GV denotes GeoVirus, and SE denotes SemEval.}

	\label{acc161all}
	\footnotesize
\begin{tabular}{ccccccccccccc}
\hline
            & LGL           & NEEL          & TN            & GWN           & GC            & GV            & WikToR        & WOTR          & CLDW         & TUD           & SE            & NCEN          \\ \hline
Fishing     & 0.53          & 0.42          & 0.60           & 0.56          & 0.47          & 0.86          & 0.35          & 0.54          & 0.60          & 0.76          & 0.58          & 0.63          \\
DCA         & 0.41          & 0.62          & 0.46          & 0.36          & 0.61          & 0.44          & 0.17          & 0.22          & 0.47         & 0.61          & 0.57          & 0.51          \\
MulRel      & 0.69          & 0.76          & 0.66          & 0.65          & 0.71          & 0.85          & 0.26          & 0.64          & 0.67         & 0.80           & 0.68          & 0.47          \\
BLINK       & 0.71          & 0.79          & 0.76          & 0.74          & 0.75          & 0.91          & 0.64          & 0.74          & 0.78         & 0.83          & 0.75          & 0.82          \\
Bootleg     & 0.74          & 0.74          & 0.74          & 0.69          & 0.69          & 0.80           & 0.65          & 0.65          & 0.66         & 0.82          & 0.73          & 0.68          \\
GENRE       & 0.82          & 0.83          & 0.84          & 0.79          & 0.81          & 0.92          & 0.81          & 0.77          & 0.69         & 0.86          & 0.84          & 0.82          \\
ExtEnD      & 0.67          & 0.75          & 0.71          & 0.67          & 0.68          & 0.86          & 0.52          & 0.64          & 0.55         & 0.79          & 0.79          & 0.64          \\
LUKE        & 0.75          & 0.60           & 0.71          & 0.73          & 0.57          & 0.87          & 0.43          & 0.42          & 0.55         & 0.81          & 0.76          & 0.55          \\
Nominatim   & 0.64          & 0.81          & 0.68          & 0.65          & 0.74          & 0.80           & 0.21          & 0.52          & 0.24         & 0.75          & 0.80           & 0.70           \\
Adaptive    & 0.79          & 0.49          & 0.66          & 0.59          & 0.54          & 0.63          & 0.14          & 0.36          & 0.29         & 0.63          & 0.63          & 0.41          \\
Population  & 0.60           & 0.72          & 0.72          & 0.61          & 0.70           & 0.84          & 0.22          & 0.43          & 0.27         & 0.76          & 0.81          & 0.60          \\
CLAVIN      & 0.59          & 0.76          & 0.71          & 0.65          & 0.77          & 0.83          & 0.22          & 0.50           & 0.41         & 0.82          & 0.78          & 0.67          \\
TopoCluster & 0.63          & 0.79          & 0.67          & 0.68          & 0.71          & 0.78          & 0.24          & 0.61          & 0.72         & 0.79          & 0.75          & 0.72          \\
Mordecai    & 0.55          & 0.67          & 0.68          & 0.61          & 0.66          & 0.74          & 0.15          & 0.42          & 0.36         & 0.76          & 0.73          & 0.58          \\
CBH         & 0.67          & 0.31          & 0.77          & 0.64          & 0.36          & 0.73          & 0.43          & 0.54          & 0.59         & 0.72          & 0.72          & 0.57          \\
SHS         & 0.69          & 0.73          & 0.69          & 0.56          & 0.73          & 0.76          & 0.71          & 0.43          & 0.44         & 0.67          & 0.74          & 0.41          \\
CHF         & 0.68          & 0.73          & 0.77          & 0.64          & 0.75          & 0.80           & 0.43          & 0.52          & 0.56         & 0.73          & 0.79          & 0.49          \\
CamCoder    & 0.62          & 0.69          & 0.67          & 0.59          & 0.72          & 0.83          & 0.63          & 0.47          & 0.54         & 0.79          & 0.80           & 0.62          \\
Voting      & \textbf{0.83} & \textbf{0.87} & \textbf{0.87} & \textbf{0.81} & \textbf{0.85} & \textbf{0.95} & \textbf{0.85} & \textbf{0.81} & \textbf{0.8} & \textbf{0.89} & \textbf{0.87} & \textbf{0.87} \\ \hline
\end{tabular}
\end{table}

	\begin{table}[]
	\caption{\textit{AUC}  on the gold toponyms. TN denotes TR-News, GWN denotes GeoWebNews, GC denotes GeoCorpora, GV denotes GeoVirus, and SE denotes SemEval.}
	\label{aucall}
	\footnotesize
\begin{tabular}{ccccccccccccc}
\hline
            & LGL           & NEEL          & TN            & GWN           & GC            & GV            & WikToR        & WOTR          & CLDW          & TUD          & SE            & NCEN          \\ \hline
Fishing     & 0.5           & 0.58          & 0.45          & 0.47          & 0.58          & 0.22          & 0.58          & 0.49          & 0.44          & 0.32         & 0.53          & 0.36          \\
DCA         & 0.59          & 0.38          & 0.54          & 0.63          & 0.44          & 0.52          & 0.77          & 0.74          & 0.60           & 0.44         & 0.48          & 0.50           \\
MulRel         & 0.35          & 0.25          & 0.4           & 0.42          & 0.35          & 0.21          & 0.65          & 0.38          & 0.39          & 0.28         & 0.42          & 0.54          \\
BLINK       & 0.32          & 0.22          & 0.30           & 0.34          & 0.31          & 0.17          & 0.32          & 0.29          & 0.28          & 0.26         & 0.38          & 0.17          \\
Bootleg     & 0.32          & 0.28          & 0.33          & 0.39          & 0.39          & 0.26          & 0.31          & 0.38          & 0.39          & 0.28         & 0.40           & 0.32          \\
GENRE       & 0.25          & 0.18          & 0.23          & 0.29          & 0.27          & \textbf{0.15} & 0.19          & 0.27          & 0.35          & 0.23         & 0.31          & 0.16          \\
ExtEnD      & 0.38          & 0.27          & 0.36          & 0.40           & 0.4           & 0.21          & 0.45          & 0.41          & 0.49          & 0.30          & 0.36          & 0.36          \\
LUKE        & 0.30           & 0.41          & 0.35          & 0.35          & 0.49          & 0.20           & 0.55          & 0.6           & 0.47          & 0.29         & 0.37          & 0.45          \\
Nominatim   & 0.40           & 0.30           & 0.39          & 0.4           & 0.37          & 0.37          & 0.67          & 0.47          & 0.76          & 0.34         & 0.39          & 0.34          \\
Adaptive    & 0.23          & 0.57          & 0.33          & 0.41          & 0.47          & 0.48          & 0.85          & 0.70           & 0.73          & 0.40          & 0.38          & 0.63          \\
Population  & 0.36          & 0.37          & 0.26          & 0.37          & 0.31          & 0.30           & 0.67          & 0.56          & 0.72          & 0.30          & 0.20           & 0.45          \\
CLAVIN      & 0.36          & 0.32          & 0.26          & 0.33          & 0.23          & 0.29          & 0.67          & 0.49          & 0.58          & 0.21         & 0.23          & 0.37          \\
TopoCluster & 0.38          & 0.32          & 0.37          & 0.38          & 0.37          & 0.37          & 0.64          & 0.42          & 0.32          & 0.31         & 0.37          & 0.32          \\
Mordecai    & 0.42          & 0.40           & 0.31          & 0.38          & 0.34          & 0.38          & 0.78          & 0.57          & 0.65          & 0.28         & 0.28          & 0.46          \\
CBH         & 0.29          & 0.71          & 0.21          & 0.34          & 0.65          & 0.38          & 0.48          & 0.48          & 0.44          & 0.31         & 0.26          & 0.46          \\
SHS         & 0.29          & 0.36          & 0.29          & 0.41          & 0.28          & 0.35          & 0.30           & 0.55          & 0.56          & 0.36         & 0.25          & 0.60           \\
CHF         & 0.28          & 0.36          & 0.21          & 0.34          & 0.26          & 0.32          & 0.48          & 0.49          & 0.45          & 0.30          & 0.21          & 0.53          \\
CamCoder    & 0.33          & 0.39          & 0.31          & 0.39          & 0.27          & 0.30           & 0.34          & 0.52          & 0.47          & 0.25         & 0.20           & 0.43          \\
Voting      & \textbf{0.24} & \textbf{0.17} & \textbf{0.21} & \textbf{0.27} & \textbf{0.23} & \textbf{0.15} & \textbf{0.17} & \textbf{0.25} & \textbf{0.25} & \textbf{0.2} & \textbf{0.28} & \textbf{0.14} \\ \hline
\end{tabular}
\end{table}

	\begin{table}[]
	\caption{\textit{Mean Area (km)}  on the gold toponyms. TN denotes TR-News,  GWN denotes GeoWebNews, GC denotes GeoCorpora,GV denotes GeoVirus, and SE denotes SemEval.}
	\label{meall}
	\footnotesize
\begin{tabular}{ccccccccccccc}
\hline
            & LGL          & NEEL         & TN           & GWN          & GC           & GV          & WikToR       & WOTR         & CLDW         & TUD          & SE           & NCEN         \\ \hline
Fishing     & 5587         & 10786        & 5424         & 5583         & 9262         & 2085        & 5542         & 6998         & 6988         & 3137         & 7271         & 6582         \\
DCA         & 5672         & 5192         & 4703         & 7791         & 5238         & 4994        & 10449        & 9836         & 8635         & 3908         & 3408         & 7222         \\
MulRel         & 1536         & 2122         & 3364         & 4218         & 3225         & 1536        & 6113         & 3177         & 5722         & 1908         & 2804         & 8356         \\
BLINK       & 899          & 1067         & 1655         & 2453         & 1585         & 751         & 1913         & 1023         & 2322         & 1218         & 2484         & 1779         \\
Bootleg     & 1909         & 3942         & 2943         & 4718         & 4436         & 2688        & 2468         & 4483         & 6050         & 2308         & 3325         & 5825         \\
GENRE       & \textbf{303} & 878          & 509          & 1238         & 828          & 293         & 1813         & 475          & 4383         & 411          & 534          & 895          \\
ExtEnD      & 3243         & 3727         & 3099         & 4669         & 4630         & 1997        & 5399         & 5173         & 7939         & 2729         & 2234         & 6522         \\
LUKE        & 809          & 6899         & 3431         & 2527         & 6822         & 960         & 8094         & 9638         & 5761         & 1770         & 1888         & 8472         \\
Nominatim   & 1319         & 1093         & 1412         & 1961         & 1730         & 676         & 4244         & 2234         & 13509        & 1013         & 752          & 3719         \\
Adaptive    & 3809         & 8768         & 5842         & 7538         & 8668         & 6059        & 15844        & 11781        & 13932        & 6661         & 5989         & 11338        \\
Population  & 2364         & 2930         & 1935         & 4848         & 3681         & 874         & 4458         & 5513         & 10490        & 3087         & 1601         & 5739         \\
CLAVIN      & 2677         & 2064         & 2862         & 4646         & 2775         & 764         & 4425         & 3911         & 7262         & 1766         & 1228         & 4618         \\
TopoCluster & 1286         & 792          & 1213         & 1064         & 1035         & 571         & 4143         & 1320         & 1415         & 631          & 734          & 1308         \\
Mordecai    & 4273         & 3960         & 3600         & 5041         & 4967         & 3265        & 9641         & 6275         & 10031        & 2787         & 3313         & 6248         \\
CBH         & 1169         & 12773        & 1593         & 4083         & 11936        & 2134        & 1819         & 4014         & 5628         & 2991         & 1903         & 5913         \\
SHS         & 2021         & 2597         & 2375         & 5168         & 3151         & 1490        & 1621         & 5202         & 8133         & 3352         & 1993         & 6791         \\
CHF         & 946          & 2619         & 1285         & 3762         & 2983         & 933         & 1819         & 3971         & 5814         & 2612         & 1392         & 5977         \\
CamCoder    & 2247         & 3326         & 3022         & 4967         & 3504         & 698         & 1143         & 6049         & 6101         & 2445         & 1375         & 5322         \\
Voting     & 330          & \textbf{353} & \textbf{369} & \textbf{647} & \textbf{486} & \textbf{73} & \textbf{945} & \textbf{382} & \textbf{966} & \textbf{190} & \textbf{245} & \textbf{563} \\ \hline
\end{tabular}
\end{table}

	\begin{table}[]
	\caption{\textit{Accuracy@161km}  on the subset of gold toponyms correctly recognized by DBpedia Spotlight. TN denotes TR-News, GWN denotes GeoWebNews, GC denotes GeoCorpora, GV denotes GeoVirus, and SE denotes SemEval.}
	\label{acc161db}
	
	\footnotesize
\begin{tabular}{ccccccccccccc}
\hline
            & LGL           & NEEL          & TN            & GWN           & GC            & GV            & WikToR        & WOTR          & CLDW          & TUD           & SE            & NCEN          \\ \hline
DBpedia     & 0.63          & 0.72          & 0.73          & 0.67          & 0.72          & 0.70           & 0.30           & 0.65          & 0.90           & 0.73          & 0.67          & 0.94          \\
Fishing     & 0.60           & 0.48          & 0.68          & 0.72          & 0.61          & 0.90           & 0.40           & 0.68          & 0.87          & 0.83          & 0.63          & 0.80           \\
DCA         & 0.49          & 0.69          & 0.53          & 0.45          & 0.69          & 0.44          & 0.24          & 0.29          & 0.77          & 0.68          & 0.57          & 0.71          \\
MulRel      & 0.72          & 0.78          & 0.70           & 0.73          & 0.80           & 0.88          & 0.32          & 0.70           & 0.94          & 0.87          & 0.71          & 0.54          \\
BLINK       & 0.74          & 0.82          & 0.82          & 0.83          & 0.81          & 0.93          & 0.65          & 0.80           & 0.93          & 0.87          & 0.81          & 0.94          \\
Bootleg     & 0.79          & 0.86          & 0.83          & 0.82          & 0.82          & 0.83          & 0.71          & 0.80           & 0.94          & 0.88          & 0.82          & 0.88          \\
GENRE       & 0.82          & 0.86          & 0.88          & 0.85          & 0.84          & 0.94          & 0.82          & 0.82          & 0.88          & 0.89          & 0.85          & 0.93          \\
ExtEnD      & 0.70           & 0.83          & 0.77          & 0.78          & 0.77          & 0.9           & 0.51          & 0.76          & 0.82          & 0.86          & 0.83          & 0.81          \\
LUKE        & 0.79          & 0.71          & 0.78          & 0.82          & 0.69          & 0.89          & 0.47          & 0.61          & 0.82          & 0.88          & 0.81          & 0.73          \\
Nominatim   & 0.71          & 0.85          & 0.76          & 0.76          & 0.83          & 0.82          & 0.27          & 0.61          & 0.32          & 0.81          & 0.89          & 0.89          \\
Adaptive    & 0.86          & 0.54          & 0.69          & 0.72          & 0.70           & 0.68          & 0.16          & 0.42          & 0.44          & 0.74          & 0.66          & 0.54          \\
Population  & 0.7           & 0.79          & 0.80           & 0.79          & 0.84          & 0.87          & 0.27          & 0.61          & 0.49          & 0.88          & 0.82          & 0.81          \\
CLAVIN      & 0.71          & 0.81          & 0.80           & 0.82          & \textbf{0.91} & 0.86          & 0.27          & 0.64          & 0.69          & 0.91          & 0.84          & 0.87          \\
TopoCluster & 0.71          & 0.82          & 0.76          & 0.78          & 0.86          & 0.81          & 0.26          & 0.72          & 0.88          & 0.84          & 0.78          & 0.88          \\
Mordecai    & 0.63          & 0.74          & 0.76          & 0.77          & 0.79          & 0.80           & 0.19          & 0.51          & 0.54          & 0.84          & 0.79          & 0.73          \\
CBH         & 0.73          & 0.33          & 0.81          & 0.75          & 0.42          & 0.75          & 0.47          & 0.61          & 0.86          & 0.78          & 0.73          & 0.75          \\
SHS         & 0.73          & 0.8           & 0.70           & 0.69          & 0.85          & 0.81          & 0.72          & 0.49          & 0.7           & 0.72          & 0.76          & 0.54          \\
CHF         & 0.74          & 0.79          & 0.80           & 0.75          & 0.87          & 0.83          & 0.47          & 0.56          & 0.83          & 0.79          & 0.8           & 0.65          \\
CamCoder    & 0.73          & 0.76          & 0.77          & 0.77          & 0.84          & 0.85          & 0.67          & 0.61          & 0.81          & 0.89          & 0.82          & 0.81          \\
Voting      & \textbf{0.83} & \textbf{0.89} & \textbf{0.92} & \textbf{0.87} & 0.88          & \textbf{0.96} & \textbf{0.85} & \textbf{0.86} & \textbf{0.96} & \textbf{0.92} & \textbf{0.86} & \textbf{0.97} \\ \hline
\end{tabular}
\end{table}

	\begin{table}[]
	\caption{\textit{AUC}  on the subset of gold toponyms correctly recognized by DBpedia Spotlight. TN denotes TR-News, GWN denotes GeoWebNews, GC denotes GeoCorpora, GV denotes GeoVirus, and SE denotes SemEval.}
	\label{aucdb}
	\footnotesize
\begin{tabular}{ccccccccccccc}
\hline
            & LGL           & NEEL          & TN            & GWN           & GC           & GV            & WikToR        & WOTR          & CLDW          & TUD           & SE            & NCEN          \\ \hline
DBpedia     & 0.40           & 0.26          & 0.32          & 0.33          & 0.31         & 0.31          & 0.64          & 0.34          & 0.16          & 0.31          & 0.39          & 0.07          \\
Fishing     & 0.46          & 0.51          & 0.40           & 0.34          & 0.45         & 0.16          & 0.54          & 0.36          & 0.20           & 0.27          & 0.48          & 0.19          \\
DCA         & 0.54          & 0.30           & 0.50           & 0.55          & 0.37         & 0.50           & 0.71          & 0.63          & 0.34          & 0.38          & 0.44          & 0.31          \\
MulRel      & 0.35          & 0.21          & 0.38          & 0.35          & 0.27         & 0.17          & 0.61          & 0.29          & 0.14          & 0.24          & 0.38          & 0.48          \\
BLINK       & 0.32          & 0.17          & 0.28          & 0.27          & 0.26         & 0.14          & 0.32          & 0.23          & 0.14          & 0.23          & 0.33          & \textbf{0.05} \\
Bootleg     & 0.29          & 0.16          & 0.28          & 0.28          & 0.28         & 0.23          & 0.27          & 0.23          & 0.13          & 0.23          & 0.33          & 0.13          \\
GENRE       & 0.26          & \textbf{0.15} & 0.22          & 0.25          & 0.25         & 0.13          & 0.19          & 0.21          & 0.18          & 0.22          & 0.30           & 0.07          \\
ExtEnD      & 0.37          & 0.19          & 0.33          & 0.31          & 0.32         & 0.17          & 0.46          & 0.28          & 0.25          & 0.25          & 0.32          & 0.19          \\
LUKE        & 0.29          & 0.30           & 0.31          & 0.27          & 0.38         & 0.17          & 0.51          & 0.43          & 0.22          & 0.24          & 0.33          & 0.27          \\
Nominatim   & 0.37          & 0.27          & 0.33          & 0.33          & 0.33         & 0.36          & 0.65          & 0.40           & 0.70           & 0.31          & 0.36          & 0.17          \\
Adaptive    & \textbf{0.17} & 0.53          & 0.31          & 0.28          & 0.32         & 0.44          & 0.84          & 0.63          & 0.59          & 0.31          & 0.34          & 0.51          \\
Population  & 0.29          & 0.31          & 0.20           & 0.20           & 0.18         & 0.28          & 0.65          & 0.43          & 0.53          & 0.2           & 0.19          & 0.27          \\
CLAVIN      & 0.28          & 0.29          & \textbf{0.18} & \textbf{0.17} & \textbf{0.10} & 0.28          & 0.64          & 0.37          & 0.34          & \textbf{0.14} & \textbf{0.18} & 0.18          \\
TopoCluster & 0.35          & 0.30           & 0.32          & 0.31          & 0.29         & 0.36          & 0.64          & 0.34          & 0.18          & 0.29          & 0.36          & 0.20           \\
Mordecai    & 0.36          & 0.35          & 0.24          & 0.22          & 0.24         & 0.34          & 0.76          & 0.50           & 0.50           & 0.21          & 0.21          & 0.33          \\
CBH         & 0.25          & 0.70           & 0.19          & 0.23          & 0.60          & 0.37          & 0.46          & 0.41          & 0.19          & 0.27          & 0.26          & 0.29          \\
SHS         & 0.27          & 0.31          & 0.29          & 0.29          & 0.17         & 0.33          & 0.30           & 0.49          & 0.33          & 0.32          & 0.23          & 0.48          \\
CHF         & 0.24          & 0.31          & 0.20           & 0.23          & 0.16         & 0.31          & 0.46          & 0.44          & 0.21          & 0.26          & 0.20           & 0.38          \\
CamCoder    & 0.25          & 0.33          & 0.23          & 0.22          & 0.16         & 0.28          & 0.33          & 0.40           & 0.23          & 0.17          & 0.18          & 0.25          \\
Voting      & 0.26          & \textbf{0.15} & 0.19          & 0.23          & 0.22         & \textbf{0.12} & \textbf{0.18} & \textbf{0.19} & \textbf{0.11} & 0.2           & 0.29          & 0.06          \\ \hline
\end{tabular}
\end{table}

	\begin{table}[]
	\caption{\textit{Mean Error (km)}  on the subset of gold toponyms correctly recognized by DBpedia Spotlight. TN denotes TR-News, GWN denotes GeoWebNews, GC denotes GeoCorpora, GV denotes GeoVirus, and SE denotes SemEval.}
	\label{medb}
	\footnotesize
\begin{tabular}{cllllllllllll}
\hline
            & \multicolumn{1}{c}{LGL} & \multicolumn{1}{c}{NEEL} & \multicolumn{1}{c}{TN} & \multicolumn{1}{c}{GWN} & \multicolumn{1}{c}{GC} & \multicolumn{1}{c}{GV} & \multicolumn{1}{c}{WikToR} & \multicolumn{1}{c}{WOTR} & \multicolumn{1}{c}{CLDW} & \multicolumn{1}{c}{TUD} & \multicolumn{1}{c}{SE} & \multicolumn{1}{c}{NCEN} \\ \hline
DBpedia     & 1847                    & 648                      & 1683                   & 1310                    & 1201                   & 1044                   & 5498                       & 1629                     & 1110                     & 938                     & 335                    & 144                      \\
Fishing     & 4290                    & 9252                     & 4448                   & 2537                    & 5934                   & 1344                   & 4403                       & 4138                     & 1997                     & 1756                    & 5503                   & 3274                     \\
DCA         & 4126                    & 3082                     & 3797                   & 5271                    & 3134                   & 4219                   & 8803                       & 6093                     & 2302                     & 2039                    & 1337                   & 2474                     \\
MulRel      & 1298                    & 907                      & 3181                   & 2761                    & 998                    & 1079                   & 5206                       & 1111                     & 543                      & 535                     & 1550                   & 7171                     \\
BLINK       & 803                     & 376                      & 1446                   & 1304                    & 541                    & 559                    & 1979                       & 627                      & 519                      & 432                     & 843                    & 370                      \\
Bootleg     & 973                     & 1348                     & 1874                   & 1933                    & 1419                   & 2248                   & 1960                       & 1631                     & 723                      & 909                     & 900                    & 1968                     \\
GENRE       & \textbf{331}            & 647                      & \textbf{281}           & 654                     & 479                    & 286                    & 1671                       & 349                      & 1670                     & 227                     & 275                    & 375                      \\
ExtEnD      & 2542                    & 1833                     & 2423                   & 2360                    & 2381                   & 1326                   & 5070                       & 2634                     & 3018                     & 1240                    & 814                    & 2996                     \\
LUKE        & 581                     & 4328                     & 2623                   & 741                     & 3915                   & 629                    & 7086                       & 5763                     & 1216                     & 491                     & 690                    & 4938                     \\
Nominatim   & 1203                    & 549                      & 1090                   & 684                     & 703                    & 374                    & 4327                       & 1239                     & 13100                    & 331                     & 380                    & 732                      \\
Adaptive    & 2653                    & 7607                     & 5355                   & 4877                    & 5507                   & 5099                   & 15354                      & 10408                    & 10981                    & 4711                    & 5445                   & 8690                     \\
Population  & 1697                    & 1435                     & 1163                   & 2019                    & 1510                   & 568                    & 4492                       & 3127                     & 7665                     & 1194                    & 1429                   & 1941                     \\
CLAVIN      & 1676                    & 1068                     & 1685                   & 2045                    & 676                    & 629                    & 4459                       & 1717                     & 3805                     & 553                     & 677                    & 1093                     \\
TopoCluster & 1246                    & 649                      & 1040                   & 524                     & 427                    & 351                    & 4287                       & 910                      & 877                      & 274                     & 581                    & 739                      \\
Mordecai    & 3516                    & 2790                     & 2205                   & 2682                    & 3364                   & 2315                   & 9184                       & 4228                     & 7946                     & 1574                    & 1578                   & 3749                     \\
CBH         & 1013                    & 12256                    & 1248                   & 2131                    & 11023                  & 2094                   & 1730                       & 2673                     & 1556                     & 1918                    & 1589                   & 2260                     \\
SHS         & 1801                    & 1189                     & 2087                   & 2762                    & 1199                   & 1006                   & 1666                       & 3719                     & 4117                     & 2168                    & 1712                   & 3504                     \\
CHF         & 814                     & 1249                     & 1068                   & 1823                    & 1075                   & 710                    & 1730                       & 2833                     & 1749                     & 1480                    & 1034                   & 2413                     \\
CamCoder    & 1210                    & 1828                     & 2093                   & 2099                    & 1417                   & 562                    & 1063                       & 2682                     & 2013                     & 913                     & 1004                   & 1476                     \\
Voting      & 383                     & \textbf{327}             & 290                    & \textbf{187}            & \textbf{264}           & \textbf{37}            & \textbf{995}               & \textbf{303}             & \textbf{172}             & \textbf{100}            & \textbf{167}           & \textbf{121}             \\ \hline
\end{tabular}
\end{table}

	\begin{table}[]
	\caption{\textit{Accuracy@161km}  on the subset of gold toponyms correctly recognized by Edinburgh Geoparser. TN denotes TR-News, GWN denotes GeoWebNews, GC denotes GeoCorpora, GV denotes GeoVirus, and SE denotes SemEval.}
	\label{acced}
	\footnotesize
\begin{tabular}{ccccccccccccc}
\hline
            & LGL           & NEEL          & TN           & GWN           & GC            & GV            & WikToR        & WOTR          & CLDW          & TUD           & SE            & NCEN          \\ \hline
Edinburgh   & 0.8           & 0.50           & 0.72         & 0.89          & 0.87          & 0.69          & 0.56          & 0.73          & 0.58          & 0.91          & 0.88          & 0.80           \\
Fishing     & 0.58          & 0.45          & 0.67         & 0.70           & 0.61          & 0.88          & 0.40           & 0.61          & 0.69          & 0.81          & 0.80           & 0.79          \\
DCA         & 0.45          & 0.67          & 0.50          & 0.46          & 0.68          & 0.45          & 0.21          & 0.27          & 0.56          & 0.67          & 0.64          & 0.69          \\
MulRel      & 0.71          & 0.78          & 0.67         & 0.74          & 0.79          & 0.86          & 0.29          & 0.69          & 0.77          & 0.85          & 0.72          & 0.55          \\
BLINK       & 0.74          & 0.79          & 0.78         & 0.85          & 0.82          & 0.93          & 0.65          & 0.77          & 0.83          & 0.85          & 0.82          & 0.93          \\
Bootleg     & 0.79          & 0.79          & 0.79         & 0.83          & 0.80           & 0.82          & 0.72          & 0.73          & 0.76          & 0.87          & 0.84          & 0.84          \\
GENRE       & 0.83          & 0.85          & 0.87         & 0.89          & 0.84          & 0.94          & 0.82          & 0.81          & 0.76          & 0.88          & 0.88          & 0.90           \\
ExtEnD      & 0.72          & 0.8           & 0.74         & 0.81          & 0.77          & 0.89          & 0.55          & 0.74          & 0.63          & 0.85          & 0.85          & 0.81          \\
LUKE        & 0.77          & 0.67          & 0.73         & 0.86          & 0.69          & 0.89          & 0.44          & 0.51          & 0.62          & 0.87          & 0.80           & 0.69          \\
Nominatim   & 0.69          & 0.88          & 0.75         & 0.82          & 0.85          & 0.81          & 0.24          & 0.58          & 0.26          & 0.78          & 0.85          & 0.86          \\
Adaptive    & 0.82          & 0.58          & 0.66         & 0.75          & 0.75          & 0.68          & 0.15          & 0.43          & 0.33          & 0.73          & 0.63          & 0.55          \\
Population  & 0.66          & 0.84          & 0.78         & 0.85          & 0.85          & 0.86          & 0.24          & 0.49          & 0.31          & 0.86          & 0.80           & 0.78          \\
CLAVIN      & 0.68          & 0.85          & 0.80          & 0.90           & 0.92          & 0.85          & 0.24          & 0.56          & 0.46          & 0.91          & 0.80           & 0.84          \\
TopoCluster & 0.68          & 0.8           & 0.72         & 0.84          & 0.84          & 0.81          & 0.26          & 0.67          & 0.77          & 0.83          & 0.79          & 0.83          \\
Mordecai    & 0.6           & 0.73          & 0.75         & 0.78          & 0.76          & 0.76          & 0.16          & 0.45          & 0.39          & 0.82          & 0.75          & 0.70           \\
CBH         & 0.68          & 0.37          & 0.75         & 0.75          & 0.39          & 0.74          & 0.46          & 0.62          & 0.66          & 0.78          & 0.75          & 0.72          \\
SHS         & 0.73          & 0.81          & 0.68         & 0.72          & 0.84          & 0.80          & 0.73          & 0.51          & 0.50           & 0.75          & 0.76          & 0.52          \\
CHF         & 0.69          & 0.82          & 0.75         & 0.76          & 0.86          & 0.82          & 0.46          & 0.60           & 0.63          & 0.80           & 0.80           & 0.61          \\
CamCoder    & 0.70           & 0.79          & 0.76         & 0.84          & 0.86          & 0.86          & 0.64          & 0.56          & 0.60           & 0.89          & 0.85          & 0.79          \\
Voting      & \textbf{0.84} & \textbf{0.89} & \textbf{0.9} & \textbf{0.92} & \textbf{0.89} & \textbf{0.95} & \textbf{0.85} & \textbf{0.85} & \textbf{0.85} & \textbf{0.92} & \textbf{0.91} & \textbf{0.96} \\ \hline
\end{tabular}
\end{table}

	\begin{table}[]
	\caption{\textit{AUC} on the subset of gold toponyms correctly recognized by Edinburgh Geoparser. TN denotes TR-News, GWN denotes GeoWebNews, GC denotes GeoCorpora, GV denotes GeoVirus, and SE denotes SemEval.}
	\label{auced}
	\footnotesize
\begin{tabular}{ccccccccccccc}
\hline
            & LGL           & NEEL          & TN            & GWN           & GC           & GV            & WikToR        & WOTR          & CLDW          & TUD           & SE            & NCEN          \\ \hline
Edinburgh   & \textbf{0.19} & 0.54          & 0.38          & \textbf{0.10}  & 0.13         & 0.42          & 0.40           & 0.32          & 0.44          & 0.14          & \textbf{0.13} & 0.24          \\
Fishing     & 0.46          & 0.54          & 0.39          & 0.34          & 0.45         & 0.20           & 0.53          & 0.43          & 0.36          & 0.29          & 0.37          & 0.21          \\
DCA         & 0.55          & 0.31          & 0.51          & 0.52          & 0.38         & 0.51          & 0.72          & 0.67          & 0.52          & 0.39          & 0.42          & 0.32          \\
MulRel      & 0.33          & 0.21          & 0.39          & 0.34          & 0.28         & 0.20           & 0.62          & 0.34          & 0.29          & 0.25          & 0.4           & 0.47          \\
BLINK       & 0.3           & 0.2           & 0.29          & 0.25          & 0.26         & 0.15          & 0.31          & 0.27          & 0.23          & 0.24          & 0.34          & \textbf{0.06} \\
Bootleg     & 0.28          & 0.21          & 0.29          & 0.28          & 0.28         & 0.24          & 0.26          & 0.31          & 0.30           & 0.24          & 0.34          & 0.16          \\
GENRE       & 0.24          & \textbf{0.16} & 0.21          & 0.22          & 0.24         & 0.14          & 0.19          & 0.24          & 0.30           & 0.22          & 0.29          & 0.09          \\
ExtEnD      & 0.34          & 0.21          & 0.34          & 0.28          & 0.32         & 0.19          & 0.43          & 0.33          & 0.41          & 0.25          & 0.33          & 0.19          \\
LUKE        & 0.29          & 0.33          & 0.33          & 0.24          & 0.37         & 0.19          & 0.53          & 0.52          & 0.40           & 0.24          & 0.36          & 0.31          \\
Nominatim   & 0.37          & 0.28          & 0.34          & 0.30           & 0.33         & 0.37          & 0.66          & 0.43          & 0.74          & 0.33          & 0.39          & 0.21          \\
Adaptive    & 0.20          & 0.48          & 0.33          & 0.25          & 0.27         & 0.44          & 0.85          & 0.64          & 0.69          & 0.31          & 0.36          & 0.51          \\
Population  & 0.32          & 0.28          & 0.22          & 0.15          & 0.16         & 0.29          & 0.66          & 0.50           & 0.69          & 0.20           & 0.19          & 0.29          \\
CLAVIN      & 0.28          & 0.25          & \textbf{0.18} & \textbf{0.09} & \textbf{0.10} & 0.28          & 0.65          & 0.43          & 0.53          & \textbf{0.13} & 0.19          & 0.21          \\
TopoCluster & 0.35          & 0.32          & 0.34          & 0.29          & 0.32         & 0.37          & 0.64          & 0.37          & 0.28          & 0.30           & 0.36          & 0.23          \\
Mordecai    & 0.38          & 0.37          & 0.25          & 0.21          & 0.25         & 0.37          & 0.78          & 0.55          & 0.63          & 0.22          & 0.25          & 0.36          \\
CBH         & 0.28          & 0.67          & 0.23          & 0.23          & 0.61         & 0.37          & 0.46          & 0.40           & 0.37          & 0.26          & 0.22          & 0.33          \\
SHS         & 0.27          & 0.3           & 0.30           & 0.28          & 0.16         & 0.33          & 0.29          & 0.47          & 0.50           & 0.28          & 0.21          & 0.50           \\
CHF         & 0.27          & 0.29          & 0.23          & 0.22          & 0.15         & 0.31          & 0.46          & 0.41          & 0.39          & 0.24          & 0.18          & 0.42          \\
CamCoder    & 0.27          & 0.31          & 0.23          & 0.16          & 0.14         & 0.28          & 0.34          & 0.43          & 0.41          & 0.16          & 0.14          & 0.28          \\
Voting      & 0.23          & \textbf{0.16} & 0.20           & 0.20           & 0.21         & \textbf{0.14} & \textbf{0.17} & \textbf{0.23} & \textbf{0.21} & 0.19          & 0.27          & 0.07          \\ \hline
\end{tabular}
\end{table}

	\begin{table}[]
	\caption{\textit{Mean Error (km)}  on the subset of gold toponyms correctly recognized by Edinburgh Geoparser. TN denotes TR-News, GWN denotes GeoWebNews, GC denotes GeoCorpora, GV denotes GeoVirus, and SE denotes SemEval.}
	\label{meed}
	\footnotesize
\begin{tabular}{ccccccccccccc}
\hline
            & LGL          & NEEL         & TN           & GWN         & GC           & GV          & WikToR       & WOTR         & CLDW         & TUD         & SE           & NCEN         \\ \hline
Edinburgh   & 701          & 7782         & 1204         & 540         & 811          & 1320        & 2485         & 873          & 6118         & 356         & 295          & 1266         \\
Fishing     & 4360         & 9692         & 3817         & 2196        & 5903         & 1710        & 4415         & 5391         & 5149         & 1971        & 2968         & 3548         \\
DCA         & 4634         & 3197         & 3695         & 4507        & 3070         & 4460        & 8887         & 7461         & 6533         & 2439        & 1962         & 2761         \\
MulRel      & 1192         & 664          & 3026         & 2397        & 1059         & 1235        & 5409         & 1883         & 3649         & 746         & 2270         & 7044         \\
BLINK       & 681          & 718          & 1462         & 926         & 379          & 606         & 1787         & 713          & 1575         & 527         & 1742         & 444          \\
Bootleg     & 1088         & 2468         & 1965         & 1895        & 1551         & 2305        & 1937         & 2900         & 4007         & 1066        & 2107         & 2473         \\
GENRE       & \textbf{260} & 745          & \textbf{270} & 321         & 379          & 206         & 1687         & 313          & 3317         & 210         & 275          & 452          \\
ExtEnD      & 2445         & 2375         & 2496         & 1864        & 2198         & 1468        & 4858         & 3177         & 6137         & 1368        & 1685         & 3093         \\
LUKE        & 598          & 5140         & 2995         & 632         & 3598         & 698         & 7839         & 7608         & 4287         & 581         & 1508         & 5794         \\
Nominatim   & 1005         & 366          & 1035         & 502         & 439          & 537         & 4168         & 1377         & 13284        & 448         & 633          & 994          \\
Adaptive    & 3152         & 6266         & 5499         & 4085        & 4575         & 4843        & 15608        & 10125        & 13016        & 4699        & 6322         & 8412         \\
Population  & 1540         & 718          & 1228         & 1154        & 1078         & 587         & 4343         & 3612         & 10205        & 1204        & 1304         & 1911         \\
CLAVIN      & 1249         & 585          & 1251         & 628         & 557          & 620         & 4338         & 2311         & 6480         & 419         & 579          & 1284         \\
TopoCluster & 1039         & 620          & 1170         & 403         & 395          & 461         & 4044         & 844          & 1101         & 402         & 457          & 941          \\
Mordecai    & 3855         & 2867         & 2532         & 2331        & 3781         & 2953        & 9856         & 5603         & 9698         & 1756        & 2761         & 4050         \\
CBH         & 1027         & 11626        & 1541         & 1782        & 11097        & 2031        & 1728         & 1558         & 4287         & 1547        & 1208         & 2563         \\
SHS         & 1429         & 897          & 2172         & 2504        & 994          & 1087        & 1524         & 2718         & 7049         & 1608        & 1499         & 3625         \\
CHF         & 786          & 896          & 1361         & 1407        & 801          & 754         & 1728         & 1532         & 4495         & 1101        & 790          & 2667         \\
CamCoder    & 1092         & 1211         & 1601         & 1039        & 917          & 451         & 1102         & 3302         & 4922         & 575         & 623          & 1626         \\
Voting     & 310          & \textbf{260} & 294          & \textbf{85} & \textbf{155} & \textbf{42} & \textbf{896} & \textbf{266} & \textbf{676} & \textbf{69} & \textbf{122} & \textbf{162} \\ \hline
\end{tabular}
\end{table}


\end{document}